\documentclass[12pt, preprint]{emulateapj}
\usepackage{amsmath}
\usepackage{graphicx}

\shortauthors{L. BOCO ET AL.}
\shorttitle{BH seed formation via gaseous dynamical friction and GW emission}
\slugcomment{Accepted by ApJ}

\begin{document}

\title{Growth of supermassive black hole seeds in ETG star-forming progenitors:\\
multiple merging of stellar compact remnants via gaseous dynamical friction\\ and gravitational wave emission}

\author{L. Boco\altaffilmark{1,2,3}, A. Lapi\altaffilmark{1,2,3,4}, L. Danese\altaffilmark{1,2}}
\altaffiltext{1}{SISSA, Via Bonomea 265, 34136 Trieste, Italy}\altaffiltext{2}{IFPU-Institute for Fundamental Physics of the Universe, Via Beirut 2, 34014 Trieste, Italy}\altaffiltext{3}{INFN-Sezione di Trieste, via Valerio 2, 34127 Trieste,  Italy}\altaffiltext{4}{INAF-Osservatorio Astronomico di Trieste, via Tiepolo 11, 34131 Trieste, Italy}

\begin{abstract}
We propose a new mechanism for the growth of supermassive black hole (BH) seeds in the star-forming progenitors of local early-type galaxies (ETGs) at $z\gtrsim 1$. This envisages the migration and merging of stellar compact remnants (neutron stars and stellar-mass BHs) via gaseous dynamical friction toward the central high-density regions of such galaxies. We show that, under reasonable assumptions and initial conditions, the process can build up central BH masses of order $10^4-10^6\, M_\odot$ within some $10^7$ yr, so effectively providing heavy seeds before standard disk (Eddington-like) accretion takes over to become the dominant process for further BH growth. Remarkably, such a mechanism may provide an explanation, alternative to super-Eddington accretion rates, for the buildup of billion solar masses BHs in quasar hosts at $z\gtrsim 7$, when the age of the Universe $\lesssim 0.8$ Gyr constitutes a demanding constraint; moreover, in more common ETG progenitors at redshift $z\sim 2-6$ it can concur with disk accretion to build such large BH masses even at moderate Eddington ratios $\lesssim 0.3$ within the short star-formation duration $\lesssim$ Gyr of these systems. Finally, we investigate the perspectives to detect the merger events between the migrating stellar remnants and the accumulating central supermassive BH via gravitational wave emission with future ground and space-based detectors such as the Einstein Telescope (ET) and the Laser Interferometer Space Antenna (LISA).
\end{abstract}


\setcounter{footnote}{0}

\section{Introduction}\label{intro}

The discovery of an increasing number of active supermassive black holes (BHs) with masses $M_\bullet\gtrsim 10^9\, M_\odot$ at very high redshift $z\gtrsim 7$ in gas- and dust-rich host galaxies (e.g., Fan et al. 2006; Mortlock et al. 2011; Banados et al. 2018; Venemans et al. 2017a,b, 2018), when the age of the Universe was shorter than $\lesssim 0.8$ Gyr, rekindles the longstanding issue on how these huge masses can be accumulated within such short timescales. If the increase in BH mass is mainly driven by gas disk (Eddington-like) accretion, the characteristic $e-$folding timescale $\tau_{\rm ef}$ for the BH exponential mass growth $M_\bullet(\tau)\propto e^{\tau/\tau_{\rm ef}}$ amounts to
\begin{equation}
\tau_{\rm ef} = \frac{\eta}{(1-\eta)\, \lambda}\, t_{\rm Edd} \approx \frac{4.5\times 10^{7}}{\lambda}\, {\rm yr}~;
\end{equation}
here $\lambda\equiv L/L_{\rm Edd}$ is the Eddington ratio, i.e. the ratio between the actual to the Eddington luminosity $L_{\rm Edd}\approx 1.4\times 10^{38}\, M_\bullet/M_\odot$ erg s$^{-1}$, while $t_{\rm Edd}=M_\bullet\,c^2/L_{\rm Edd}\approx 0.4$ Gyr is the Eddington timescale, and $\eta\equiv L/\dot M_\bullet\, c^2$ is the radiative efficiency that in the last equality of the equation above has been set to the reference value $\eta\sim 10\%$ appropriate for a thin disk (Shakura \& Sunyaev 1973). If the BH featured light seed masses of order $10^2\, M_\odot$ as expected from an early generation of stars (e.g., Bromm \& Larson 2004; Greif et al. 2010; Hirano et al. 2014) and the Eddington ratios were close to $\lambda\sim 1$, a number of $\gtrsim 17\, \tau_{\rm ef}\gtrsim 0.75$ Gyr would be required to grow the BH to the measured few billion solar masses, which is critically close to the age of the Universe at the observation redshifts $z\gtrsim 7$.

In order to relieve this possible tension, two main classes of solutions have been proposed in the literature. The first invokes a super-Eddington accretion rates (e.g., Li 2012; Madau et al. 2014; Aversa et al. 2015; Volonteri et al. 2015; Lupi et al. 2016; Regan et al. 2019); even with moderately slim-disk conditions allowing $\lambda\sim $ a few, the radiative efficiency $\eta$ can get substantially reduced to values of a few percent (almost independently of the BH spin), shortening the $e-$folding time to appreciably less than $10^7$ yr. The second wayout involves mechanisms able to rapidly produce heavier BH seeds $\gtrsim 10^3-10^5\, M_\odot$, so reducing somewhat the time required to attain the final billion solar masses by standard Eddington accretion (see Mayer \& Bonoli 2019 for a recent review). One of the most appealing scenario envisages the rapid formation of BH seeds via direct collapse of gas and dust clouds within a protogalaxy, possibly induced by galaxy mergers or enhanced matter inflow along cosmic filaments (e.g., Mayer et al. 2010, 2015; Di Matteo et al. 2012, 2017). Alternatively, formation of heavy BH seeds may be driven by the efficient merging of stars inside globular clusters (e.g., Portegies Zwart et al. 2004; Devecchi et al. 2012; Latif \& Ferrara 2016), though so far no intermediate mass BH has been clearly detected at the center of local stellar systems.

The issue is also of some relevance at lower redshifts $z\sim 2-7$. This is because in the local Universe the most massive relic BHs with $M_\bullet\gtrsim$ several $10^8-10^9\, M_\odot$ are typically hosted in massive galaxies with bulge mass $M_\star\gtrsim 10^{11}\, M_\odot$ (e.g., McConnell \& Ma 2013; Kormendy \& Ho 2013), and there are extreme instances in brighter cluster galaxies where the BH mass can even exceed $M_\bullet \sim 10^{10}\, M_\odot$ (e.g., Mehrgan et al. 2019). Given that the
hosts of these monsters are early-type galaxies (ETGs; e.g., Moffett et al. 2016) most of their old stellar component must have been accumulated during a quite short main star formation episode lasting some $10^8$ yr at $z\gtrsim 1$, as demonstrated by astro-archeological measurements of their stellar ages and $\alpha$-enhanced metal content (see Thomas et al. 2005, 2010; Gallazzi et al. 2006, 2014; Johansson et al. 2012). Moreover, the well-established correlations between BH and galaxy properties (e.g., McConnell \& Ma 2013; Kormendy \& Ho 2013; Shankar et al. 2016) and the parallel evolution of the cosmic star formation rate (SFR) density for galaxies and of the luminosity density for bright quasars (e.g., Madau \& Dickinson 2014; Aird et al. 2015; Kulkarni et al. 2019) strongly suggest that the BH and stellar mass must be accumulated over comparably short timescales, thought to be ultimately determined by the energy feedback from the BH itself (see Silk \& Rees 1998; Fabian 1999; King 2005; Lapi et al. 2006, 2014; for a recent review, see King \& Pounds 2015). To grow billions solar masses in some $10^8$ yr is somewhat challenging if disk accretion starts from a light seed $\sim 10^2\, M_\odot$ and proceeds with the typical Eddington ratios $\lambda\lesssim 0.3$ estimated out to $z\lesssim 4$ (see Vestergaard \& Osmer 2009; Kelly \& Shen 2013; Vestergaard 2019); as a matter of fact, an heavy seed may help in speeding up the BH growth and in explaining huge masses $M_\bullet\gtrsim 10^9\, M_\odot$ accumulated over short timescales $\lesssim$ Gyr even at these intermediate redshifts.

In this paper we submit a new scenario to form heavy BH seeds, alternative or at least complementary to the aforementioned mechanisms, and suggest a way to test it via future gravitational wave (GW) observations. Specifically, we propose BH seeds to be formed in the inner, gas-rich regions of ETG progenitors via multiple mergers of stellar compact remnants, that can be driven to sink toward the centre by gaseous dynamical friction. The idea was inspired by a wealth of recent observational evidences concerning the population of ETG progenitors; this has been discovered thanks to wide-area far-IR/sub-mm/radio surveys and shown to be responsible for the bulk of the cosmic star formation history out to $z\lesssim 6$ (e.g., Lapi et al. 2011; Gruppioni et al. 2013, 2015; Weiss et al. 2013; Koprowski et al. 2014, 2016; Strandet et al. 2016; Novak et al. 2017; Riechers et al. 2017; Schreiber et al. 2018; Zavala et al. 2018; Wang et al. 2019). Interferometric, high-resolution observations with ALMA have allowed to reveal in these galaxies large SFRs $\gtrsim 10^2-10^3\, M_\odot$ yr$^{-1}$, considerable dust amounts $\gtrsim 10^8-10^9\, M_\odot$ and huge molecular gas reservoirs $10^{10}-10^{11}\, M_\odot$ within a central compact region of a few kiloparsecs (e.g., Scoville et al. 2014, 2016; Ikarashi et al. 2015; Simpson et al. 2015; Barro et al. 2016; Spilker et al. 2016; Tadaki et al. 2017a,b, 2018; Talia et al. 2018; Lang et al. 2019). Ensuing optical/near-IR/mid-IR followup measurements and broadband SED modeling have highlighted that these objects already comprise large stellar masses $M_\star\gtrsim 10^{11}\, M_\odot$, implying typical star-formation timescales $\tau_\psi\sim$ a few to several $10^8$ yr, as also inferred from the so called galaxy main sequence (e.g., Elbaz et al. 2007; Rodighiero et al. 2011, 2015; Speagle et al. 2014; Popesso et al. 2019; Boogaard et al. 2019; Wang et al. 2019). Finally, targeted X-ray observations have started to reveal the early growth of a supermassive BH by disk accretion in their nuclear regions, before it attains a high enough mass and power to manifest as a quasar, and to likely quench star formation and evacuate gas and dust from the host (e.g., Mullaney et al. 2012; Page et al. 2012; Delvecchio et al. 2015; Rodighiero et al. 2015, 2019; Stanley et al. 2015, 2017; Massardi et al. 2018).

\begin{figure*}
\centering
\includegraphics[width=0.65\textwidth]{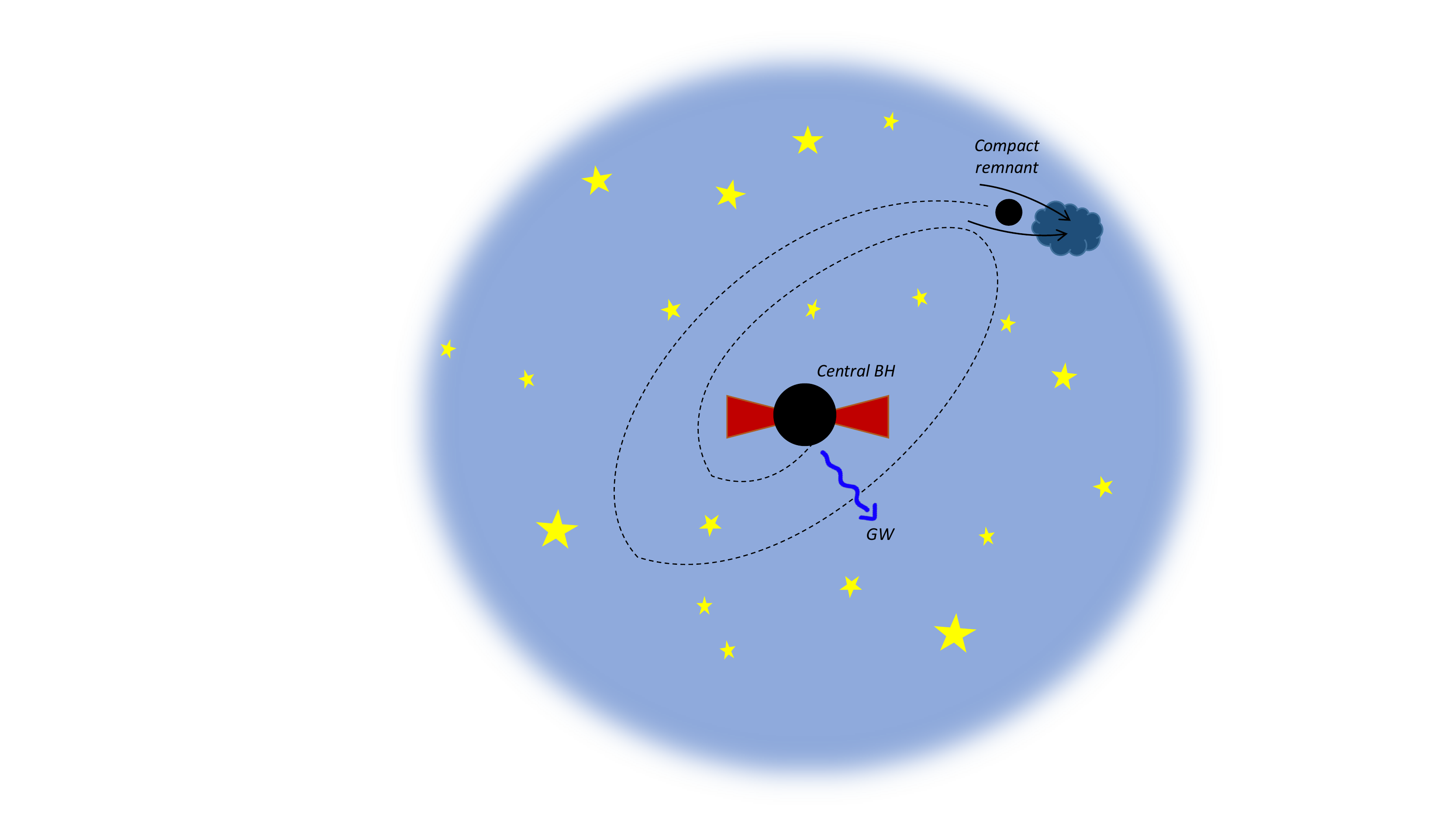}
\caption{Schematics (not to scale) depicting the migration of a stellar compact remnant due to gaseous dynamical friction toward the galaxy center, and its merging with the central BH mass (possibly also accreting matter via disk Eddington-like accretion) with ensuing emission of GWs.}
\label{fig:cartoon}
\end{figure*}

Such observational evidences reveal that in the nuclear regions of ETG progenitors a considerable amount of stars, and consequently of stellar compact remnants (neutron stars and BHs), is being formed rapidly in a very dense gaseous environment; we will show that such conditions are apt for efficient gaseous dynamical friction to occur and to drive the sinking of compact remnants toward the nuclear regions (see schematics in Fig.~\ref{fig:cartoon}). Specifically, in the present paper we will try to address the following issues: what are the typical timescales of the gaseous dynamical friction process? How this process may concur with standard disk accretion in providing an heavy BH seed and in growing the central supermassive BH? Is it possible to test this scenario via the detection of GWs emitted via the merger events between the migrating compact remnants and the accumulating central BH mass? If so, what are the marking features of this GW emission with respect to that coming from the compact binary mergers already detected by the AdvLIGO/Virgo team?

The plan of the paper is as follows: in Sect.~\ref{dynamical efficiency} we discuss the timescales of gaseous dynamical friction in driving stellar compact remnants towards the center of ETG progenitors; in Sect.~\ref{central mass} we compute the ensuing merger rates of the compact remnants and the induced time evolution of the central BH mass via dynamical friction and disk (Eddington-like) accretion; in Sect.~\ref{GW} we discuss the GW emission associated to the process of seed formation proposed here, and its detectability with future ground-based (Einstein Telescope, ET) and space-based (Laser Interferometric Space Antenna, LISA) detectors; in Sect.~\ref{discussion} we critically discuss the main assumption of our treatment; finally, in Sect.~\ref{summary} we summarize our findings and possible future developments.

\section{Gaseous dynamical friction in star-forming ETG progenitors}\label{dynamical efficiency}

Generally speaking, dynamical friction consists in the gravitational interaction between a moving object (dubbed perturber) and its gravitationally-induced wake, which generates a reduction in the energy and angular momentum of the perturber, and hence its progressive orbital decay. In the literature more emphasis has been given to the dynamical friction process against a sea of background stars or collisionless dark matter (e.g., Chandrasekhar 1943; Binney \& Tremaine 1987; Lacey \& Cole 1993; Hashimoto et al. 2003; Fujii et al. 2006; Boylan-Kolchin et al. 2008; Jiang et al. 2008). For example, this is a leading mechanism thought to drive the formation of a supermassive BH binary after a galaxy merger (see Begelman et al. 1980; Mayer et al. 2007; Barausse 2012; Chapon et al. 2013; Antonini et al. 2015; Tamburello et al. 2017; Katz et al. 2019); the binary can eventually coalesce and emit GWs if stalling around the hardening radius (the so called `final parsec problem') is avoided by some mechanism like gas dynamics, triple BH interactions, circum-nuclear disk migration, etc. (see Yu 2002; Escala et al. 2004; Merritt \& Milosavljevic 2005; Kulkarni \& Loeb 2012; Bonetti et al. 2019).

In our context of driving a stellar compact remnants to the center in  gas-rich ETG progenitors, dynamical friction against collisionless matter is of minor relevance. This is because we are mainly interested in the buildup of an heavy  BH seed before standard disk (Eddington-like) accretion becomes the dominant channel for the hole growth. At these early stages, an ETG progenitor is still poor in stellar content though extremely rich in molecular gas; moreover, such gas reservoir is expected to strongly dominate the inner gravitational potential (see next Section). Therefore gaseous rather than stellar or dark matter dynamical friction should constitute the relevant process to drive the compact remnants toward the nucleus, prevent stalling, and enforce coalescence with the accumulating central BH mass. To estimate the efficiency and timescale of the process, we need to model three basic ingredients: (i) the number density and velocity distributions of stellar compact remnants in the central regions of an ETG progenitor; (ii) the dynamical friction force acting on a stellar compact remnant during its orbit in the galactic potential well; (iii) the accretion of gas onto the stellar compact remnant during the orbital decay. These will be now discussed in turn.

\subsection{Number density and velocity distributions of compact remnants}\label{galaxy}

Resolved interferometric observations of ETG progenitors (see references in Sect.\ref{intro}) show that these objects feature a central region of size $\sim$ kpc containing huge gas masses $\gtrsim$ some $10^{10}\, M_\odot$ and undergoing large star formation at rates $\psi \gtrsim 10^2-10^3\, M_\odot$ yr$^{-1}$; these SFRs will lead to accumulate stellar masses $M_\star\gtrsim 10^{10}\, M_\odot$ over a timescale of some $10^8$ yr.
The molecular gas mass is typically found to be distributed like a Sersic profile with index $n\sim 1.5$ and half-mass radius $R_e\sim$ kpc, strongly dominating the inner gravitational potential well (the dark matter contribution is negligible out to a few tens kpcs; see van Dokkum et al. 2015; Genzel et al. 2017; Teklu et al. 2018). On such an observational basis, we adopt a $3-$D Sersic gas distribution
\begin{equation}
\rho(r)=\frac{M_{\rm gas}}{4\pi\,R_e^3}\,\frac{b^{n\,(3-\alpha)}}{n\,\Gamma[n\,(3-\alpha)]}\,\left(\frac{r}{R_e}\right)^{-\alpha}\,e^{-b\,\left(r/R_e\right)^{1/n}}
\label{density}
\end{equation}
where $R_e$ is the half-mass radius, $n$ is the Sersic index, and $\alpha$ is the inner density slope. In the classic 3-D Sersic profile (see Prugniel \& Simien 1997) $\alpha=1-1.188/2n+0.22/4n^2$ is related to $n$, yielding $\alpha\approx 0.6$ for $n=1.5$ that we adopt as our fiducial case; however, in the nuclear region, $\alpha$ can deviate somewhat from this value due to the local environment, so we will explore the impact on our results of freely varying this parameter. The corresponding mass distribution writes
\begin{equation}
M(<r)=M_{\rm gas}\,\left\{1-\frac{\Gamma[n\,(3-\alpha), b\,(r/R_e)^{1/n}]}{\Gamma[n\,(3-\alpha)]}\right\}
\label{mass}
\end{equation}
in terms of the incomplete Gamma function $\Gamma(t,a)\equiv\int_a^\infty {\rm d} t\, t^{x-1}\,e^{-t}$; the parameter $b$ can be determined numerically by the consistency condition $M_{\rm gas}(<R_e)=M_{\rm gas}/2$, which readily implies the equation $\Gamma[n\,(3-\alpha),b]=\Gamma[n\,(3-\alpha)]/2$. Finally, the associated gravitational potential is given by
\begin{equation}
\begin{aligned}
\phi(r) &=-\frac{G\, M_{\rm gas}}{R_e}\left\{\frac{1}{r}-\frac{\Gamma[n\,(3-\alpha),b\,(r/R_e)^{1/n}])}{\Gamma[n\,(3-\alpha)]}\right. +\\
\\
& \left.+b^n\,\frac{\Gamma[n\,(2-\alpha),b\, (r/R_e)^{1/n}]}{\Gamma[n\,(3-\alpha)]}\right\}.\\
\end{aligned}
\end{equation}

For comparison with previous works, we will also explore other two classic density distributions: (i) the singular isothermal sphere or SIS model, for which $\rho(r)=(M_{\rm gas}/2\pi R_e^3)\, (r/R_e)^{-2}$, $M(<r)=M_{\rm gas}\,r/2R_e$ and $\phi(r)= (G M_{\rm gas}/2R_e)\, [\log(r/2 R_e)-1]$; (ii) the Hernquist (1990) profile for which $\rho(r)=(M_{\rm gas}/2\pi R_e^3)\, (\sqrt{2}-1)\,(r/R_e)^{-1}\,(\sqrt{2}-1+r/R_e)^{-3}$, $M(<r)=M_{\rm gas}\,(r/R_e)^2\,(\sqrt{2}-1+r/R_e)^{-2}$, and $\phi(r)=-(G M_{\rm gas}/R_e)\,(\sqrt{2}-1+r/R_e)^{-1}$.

We assume stars, and hence stellar compact remnants, to be created following the above gas distribution; specifically, we prescribe that
\begin{equation}
\frac{{\rm d} p}{{\rm d} r}\propto\frac{{\rm d} M(<r)}{{\rm d} r}\propto r^2\rho(r)
\label{dpdr}
\end{equation}
is the probability that a star was born at a radius $r$. After a timescale $\sim$ a few $10^7$ yr stars more massive than $m_\star\gtrsim 7-8\,M_\odot$ will explode as supernovae leaving a compact remnant, i.e. a neutron star or a stellar-mass BH. We assume that the compact remnant inherits the same velocity of the progenitor's star, in turn being related to that of the star-forming molecular gas cloud; in particular, we take the distributions of radial and tangential velocities
\begin{equation}
\frac{{\rm d}p}{{\rm d}v_{r,\theta}}(v_{r,\theta}|r)\propto e^{-v_{r,\theta}^2/2\,\sigma^2}
\label{dpdv}
\end{equation}
to be Gaussians with null mean and a dispersion equal to the isotropic velocity dispersion at the radius $r$
\begin{equation}
\sigma^2(r)=\frac{1}{\rho(r)}\int_r^\infty{\rm d} r'\frac{\rho(r')\,M(<r')}{r'^2}~,
\label{sigma}
\end{equation}
found by self-consistently solving the isotropic Jeans equation in the aforementioned potential well. To provide some definite values useful in the sequel, consider that for $R_e\sim 1$ kpc and $M_{\rm gas}\sim 10^{11}\, M_\odot$, one finds $\sigma(r)\approx 150-300$ km s$^{-1}$ for initial radii $r\sim 10-100$ pc.

These prescriptions are used to initialize the position and velocity of the compact remnants that, in turn, determine their initial energy and angular momentum, needed for computing the dynamical friction timescales as detailed below.

\subsection{Gaseous dynamical friction force}\label{dynfriction}

Dynamical friction of massive perturbers in a smooth gaseous medium has been extensively investigated in a series of classic literature works (e.g., Dokuchaev 1964; Ruderman \& Spiegel 1971; Bisnovatyi-Kogan et al. 1979; Rephaeli \& Salpeter 1980; Ostriker 1999). These concurrently found that, when the motion of the perturber is supersonic, gaseous dynamical friction is as efficient as that occurring in a collisionless medium; contrariwise, when the motion of the perturber is subsonic, the gaseous dynamical friction gets strongly suppressed. All in all, the gaseous dynamical friction force $F_{\rm DF}$ can be generally described by the expression
\begin{equation}
F_{\rm DF}=-\frac{4\pi\,G^2\,m_\bullet^2\,\rho}{v^2}\, f(\mathcal{M})~,
\label{force}
\end{equation}
where $m_\bullet$ is the mass of the perturber, $v$ is its velocity, and $f(\mathcal{M})$ is a function of the Mach number $\mathcal{M}\equiv v/c_s$, namely the ratio of the perturber velocity to the sound speed $c_s$ of the background medium; the latter in turn can be related $c_s\equiv \sqrt{\gamma\, k_B\, T/\mu\, m_p}$ to the gas temperature $T$ in terms of the Boltzmann constant $k_B$, of the mean molecular weight $\mu\sim 0.6$ and of the adiabatic index $1\la \gamma\la 5/3$. In the environment of a gas-rich ETG progenitor, the typical temperatures of the molecular gas are found to be around $\lesssim 10-100$ K, as estimated from the far-IR/sub-mm observations of the dust emission, that is in rough thermal equilibrium with the gas (e.g., Silva et al. 1998; Pearson et al. 2013; Casey et al. 2014; da Cunha et al. 2015; Boquien et al. 2019); these values yield modest sound speeds $c_s\sim 0.3-3$ km s$^{-1}$ and, given the initial velocity distributions discussed in Sect.~\ref{galaxy}, strongly supersonic motions with $\mathcal{M}\gtrsim 10^2$ apply for the majority of the compact remnants, at least for most of their orbital evolution.

For point-like perturbers, Ostriker (1999) derived the approximate expression
\begin{equation}
f(\mathcal{M})=\left\{
\begin{aligned}
&\frac{1}{2}\ln{\left(\frac{1+\mathcal{M}}{1-\mathcal{M}}\right)}-\mathcal{M} & \mathcal{M}\leq 1~,\\
\\
& \frac{1}{2}\ln{\left(1-\frac{1}{\mathcal{M}^2}\right)}+\ln\Lambda & \mathcal{M}> 1~;
\end{aligned}
\right.
\label{subsup}
\end{equation}
here $\ln\Lambda\equiv\ln{\left(r_{\rm max}/r_{\rm min}\right)}$ is the so called Coulomb logarithm, defined in terms of the maximum and minimum `impact' parameters $r_{\rm max}$ and $r_{\rm min}$; such a shape has been numerically confirmed also for extended perturbers by Sanchez-Salcedo \& Brandenburg (2001). We base on the results of more recent numerical experiments (see Escala et al. 2004; Tanaka \& Haiman 2009; Tagawa et al. 2016) that refined the above expression, yielding
\begin{equation}
f(\mathcal{M})=\\
\left\{
\begin{aligned}
&\frac{1}{2}\,\ln\Lambda\,\left[{\rm erf}\left(\frac{\mathcal{M}}{\sqrt{2}}\right)-\sqrt{\frac{2}{\pi}}\,\mathcal{M}\,e^{-\mathcal{M}^2/2}\right]\\
& 0\leq\mathcal{M}\leq 0.8&\\
\\
&\frac{3}{2}\,\ln\Lambda\,\left[{\rm erf}\left(\frac{\mathcal{M}}{\sqrt{2}}\right)-\sqrt{\frac{2}{\pi}}\,\mathcal{M}\,e^{-\mathcal{M}^2/2}\right]\\
& 0.8\leq\mathcal{M}\leq\mathcal{M}_{\rm eq}&\\
\\
&\frac{1}{2}\,\ln{\left(1-\frac{1}{\mathcal{M}^2}\right)}+\ln\Lambda &\\
& \mathcal{M}>\mathcal{M}_{\rm eq}&
\end{aligned}
\right.
\end{equation}
where $\mathcal{M}_{\rm eq}$ is set so that $f(\mathcal{M})$ is a continuous function; we stress again that for most of the perturber's orbital evolution the $\mathcal{M}>\mathcal{M}_{\rm eq}$ case is relevant.

\begin{figure*}
\centering
\includegraphics[width=0.5\textwidth]{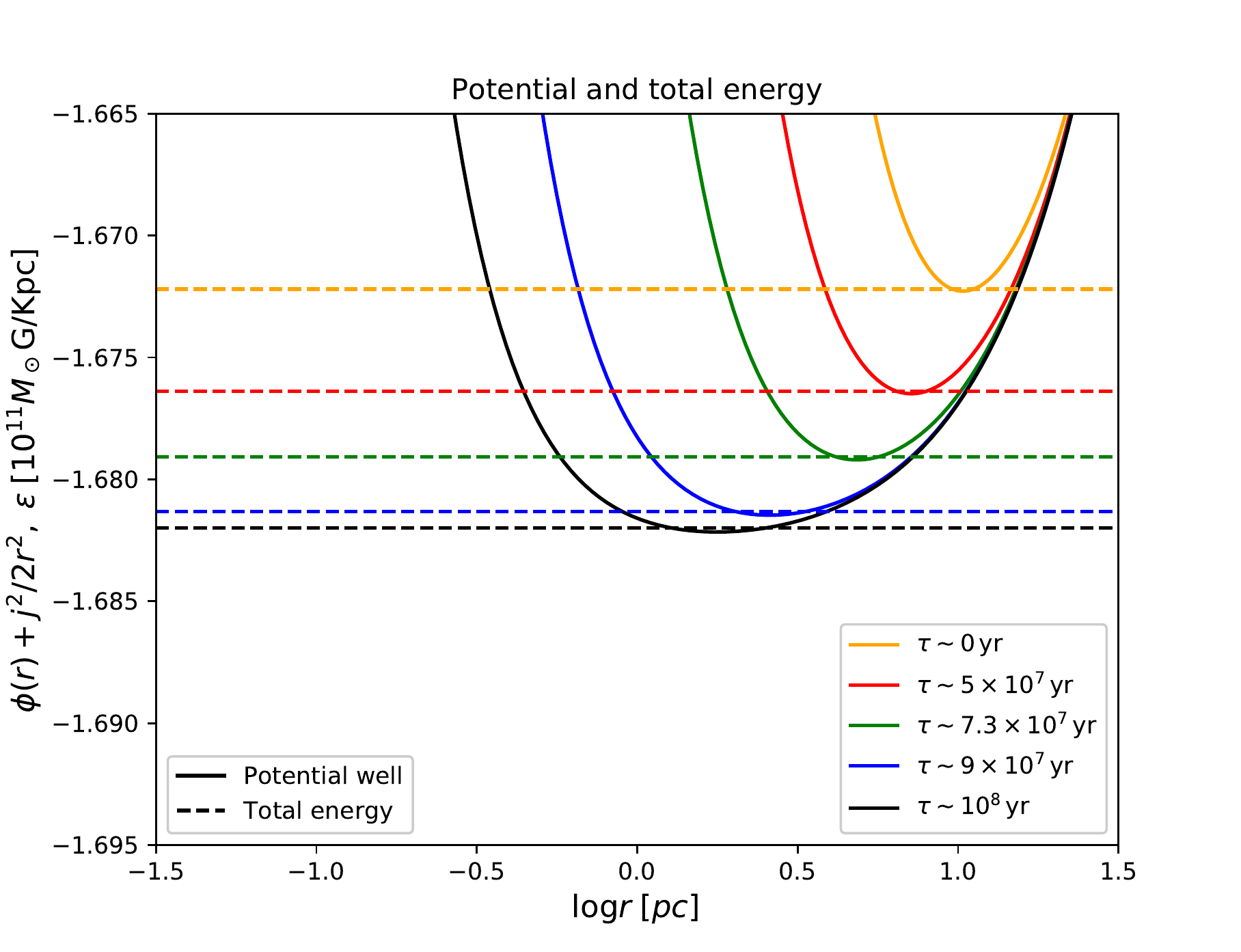}\includegraphics[width=0.5\textwidth]{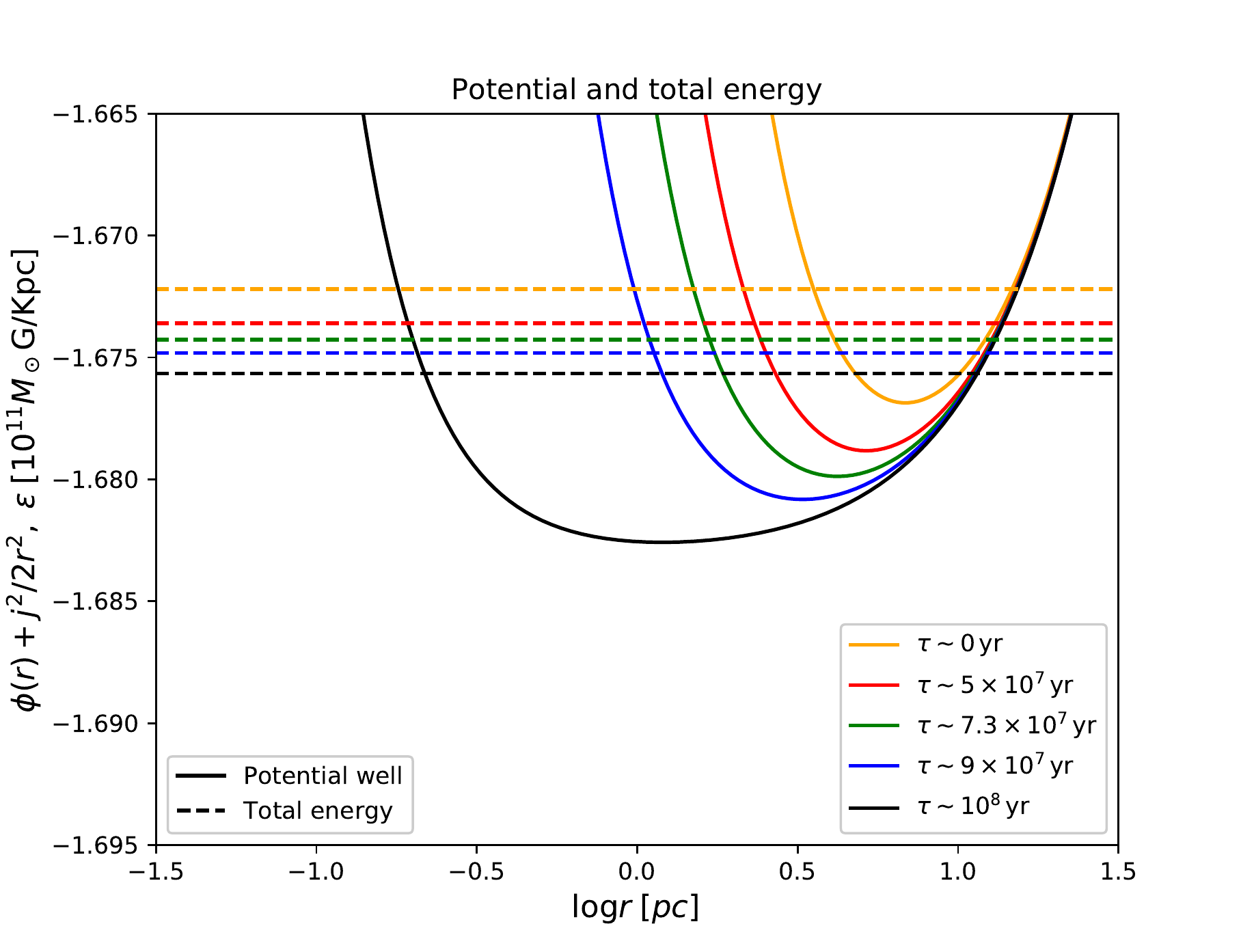}\\
\includegraphics[width=0.5\textwidth]{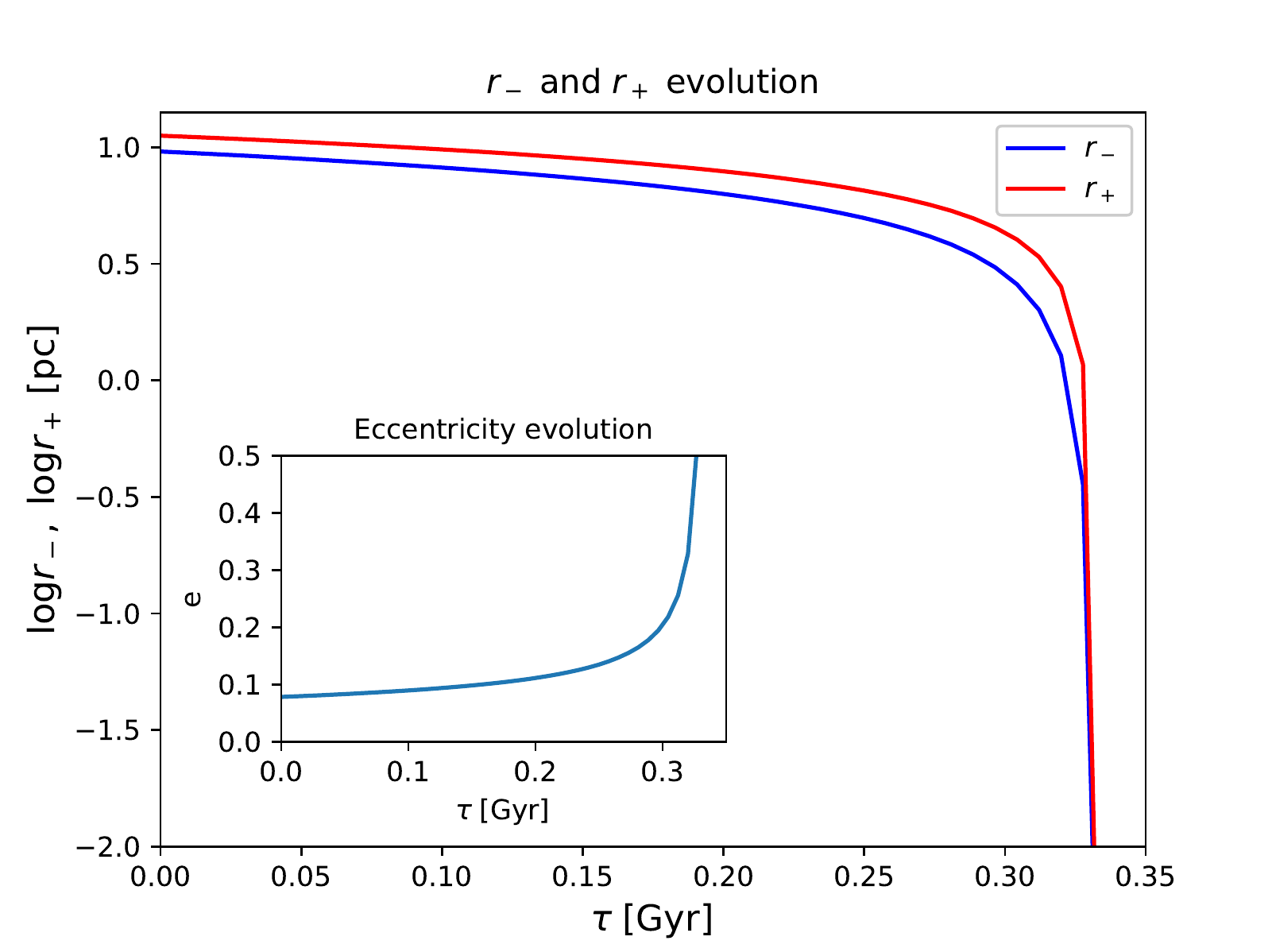}\includegraphics[width=0.5\textwidth]{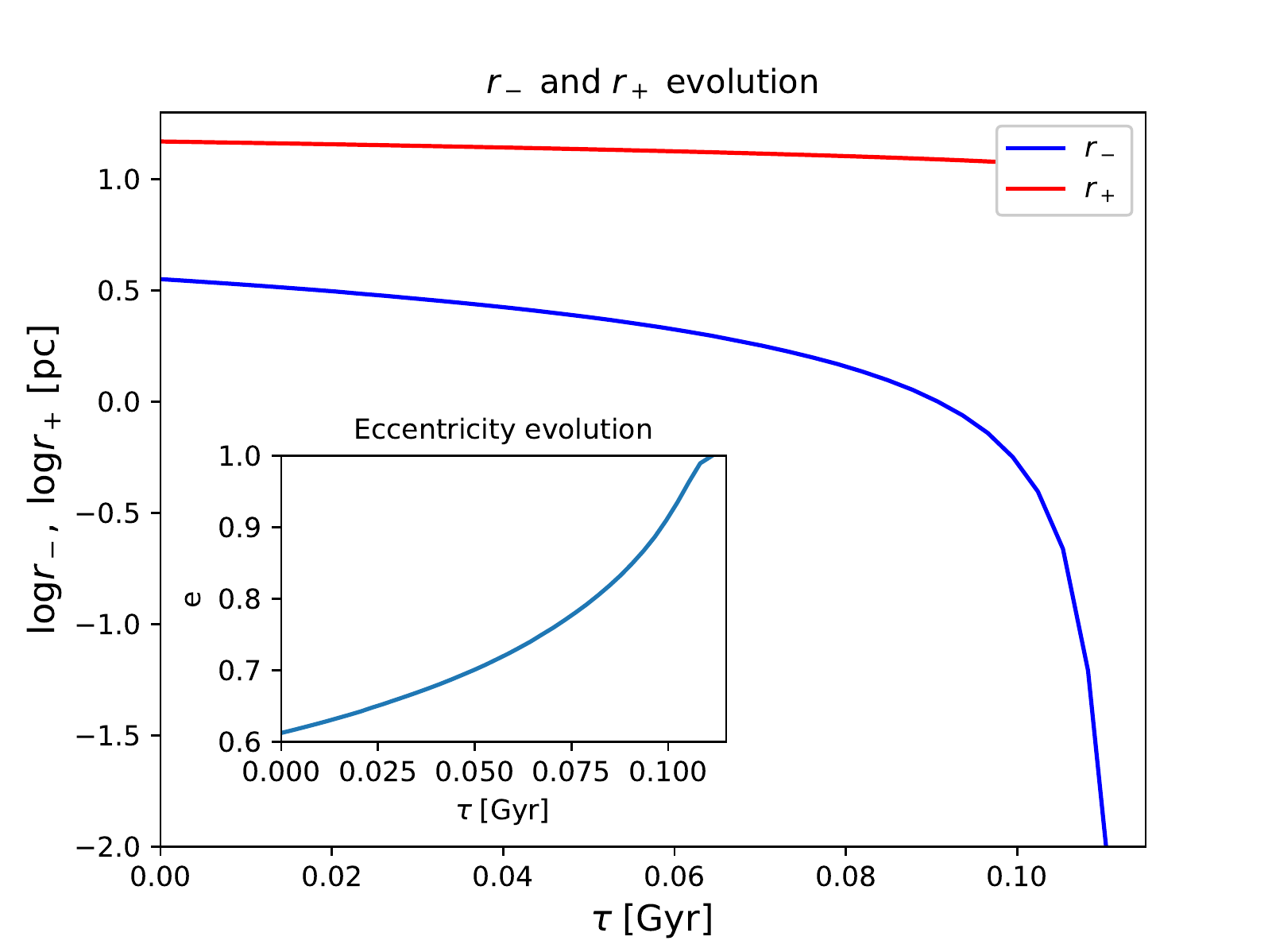}\\
\caption{Top panels: effective potential $\phi(r)+j^2/2\, r^2$ (solid lines) and total energy $\varepsilon$ (dashed lines) at different times $\tau$ (color coded as in legend) for a perturber of $m_\bullet=100\, M_\odot$ experiencing dynamical friction against a gaseous medium of mass $M_{\rm gas}=10^{11}\, M_\odot$ distributed like a Sersic profile with index $n=1.5$ and half-mass radius $R_e=1$ kpc; the initial configuration of the perturber is such that the circularity $j/j_c(\varepsilon)$ amounts to $1$ (nearly circular orbit) in the left panels and to $0.5$ (mildly eccentric orbit) in the right panels, with the same total energy. Bottom panels: evolution of the pericenter $r_-$ (blue line), of the apocenter $r_+$ (red line) and of the eccentricity $e$ (cyan line in the inset), for the same configurations as above.}
\label{fig:dynfric}
\end{figure*}

\begin{figure*}
    \centering
    \includegraphics[width=0.65\textwidth]{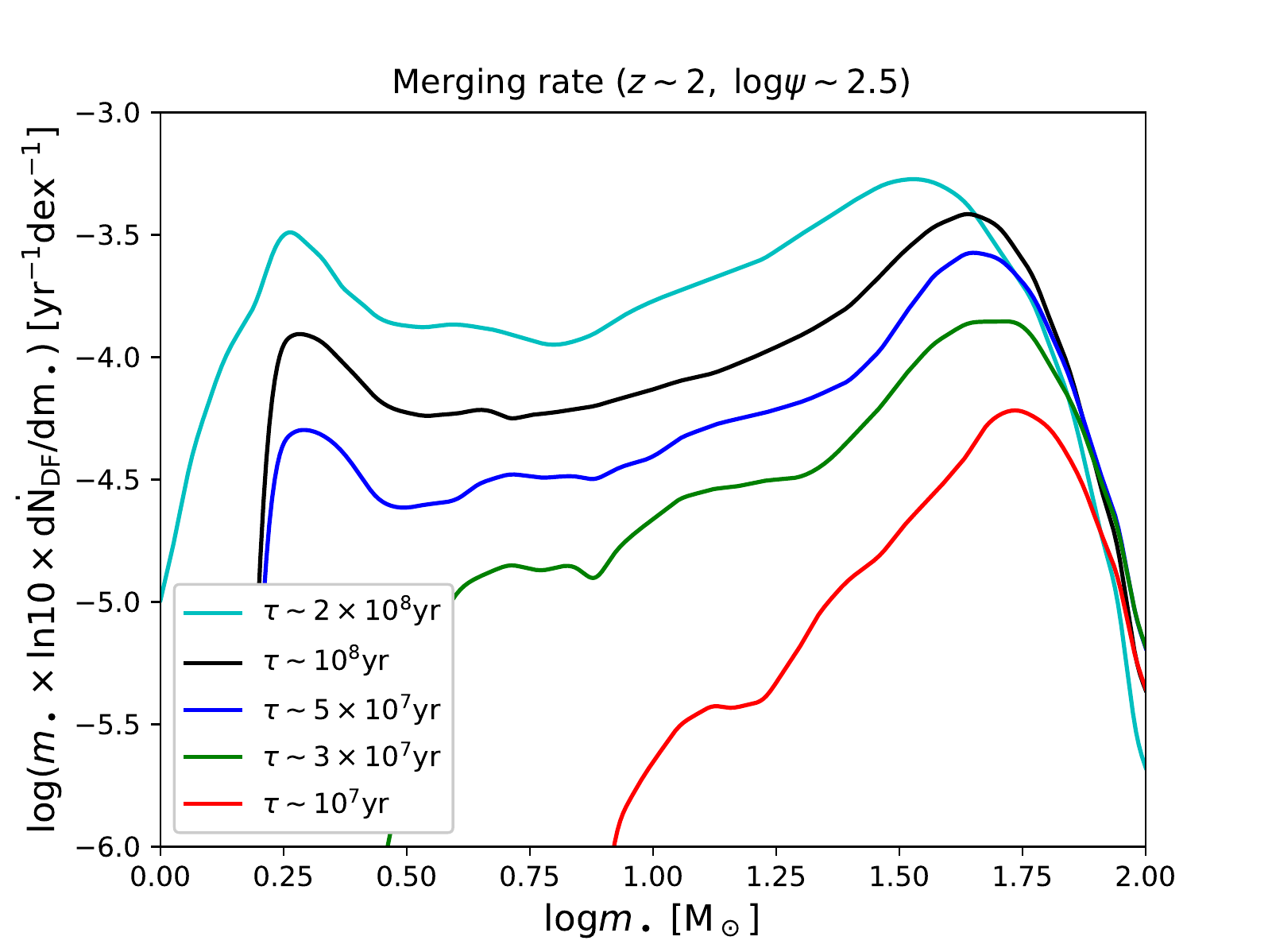}
    \caption{Merger rate due to gaseous dynamical friction per unit logarithmic bin of compact remnant mass at different galactic ages (color-coded as in legend), for a typical ETG progenitor located at $z\sim 2$ and featuring a SFR $\psi\sim 300\, M_\odot$ yr$^{-1}$.}
    \label{fig:mergerrate}
\end{figure*}

A subtle issue concerns the values of the Coulomb logarithm $\ln\Lambda\equiv \ln(r_{\rm max}/r_{\rm min})$, which brings about a considerable (though logarithmic) uncertainty for both stellar and gaseous dynamical friction. Some authors (e.g., Lacey \& Cole 1993; van den Bosch et al. 1999; Tanaka \& Haiman 2009; Tamburello et al. 2017) leave it constant during the evolution of the perturber, some others (e.g., Ostriker 1999; Tagawa et al. 2016) make it to evolve with time; moreover, the adopted values differ appreciably from author to author, though there is a general consensus for it to be $\ln\Lambda\gtrsim 1$. As to the minimum impact parameter $r_{\rm min}$, it can be identified with the accretion radius $2\, G\, m_\bullet/v^2$ if this is much larger than the softening radius of the perturber, namely the Schwartzschild radius $2\,G\, m_\bullet/c^2$ of the compact remnant in our context (see Kim \& Kim 2009; Bernal \& Sanchez-Salcedo 2013; Thun et al. 2016). The maximum impact parameter $r_{\rm max}$ is more controversial (see Binney \& Tremaine 1987), and it is often taken to be the typical scale $R_e$ of the gas distribution in which the perturber is moving (e.g., Rephaeli \& Salpeter 1980; Lacey \& Cole 1993; Silva 2016)\footnote{As a specific example, Lacey \& Cole (1993) considered the dynamical friction force on perturbers orbiting in a SIS gravitational potential of collisionless matter; they choose $\ln\Lambda=\ln (v^2\,M_{\rm tot}/V_c^2\,m_\bullet)$ where $V_c=\sqrt{G M_{\rm tot}/R_e}$ is the circular velocity and $M_{\rm tot}$ the total mass; so their prescription is formally equivalent to take $r_{\rm min}\approx G\, m_\bullet/v^2$ and $r_{\rm max}\approx R_{e}$.}; other authors commonly assume $r_{\rm max}= v\,t$ that for a straight motion (or equivalently highly eccentric orbits) would correspond to the length of the wake behind the perturber (e.g., Ostriker 1999; Tagawa et al. 2016), or a direct proportionality $r_{\rm max}=2\, r$ to the orbital radius $r$ for perturbers in nearly circular and supersonic motion (Kim \& Kim 2007).

Given this spectrum of possible choices, in this work we will explore the effect of three different prescriptions. The first one, inspired by Lacey \& Cole (1992), is to adopt $r_{\rm max}=R_e$ and $r_{\rm min}=G\, m_\bullet/v^2$ in terms of the initial velocity $v$ and mass $m_\bullet$ of the perturber, yielding a constant Coulomb logarithm $\ln\Lambda=\ln [R_{e}\,v^2/G\,m_\bullet]$. The second is to maintain the expression $\ln\Lambda=\ln [R_{e}\,v(t)^2/G\,m_\bullet(t)]$ but to use in it the running velocity $v(t)$ and mass $m_\bullet(t)$ of the perturber; the velocity changes along the orbit and on the average tends to decrease due to dynamical friction, while the mass can increase due to accretion of diffuse gas during the orbital evolution (see next Section). The third prescription, which will actually constitute our fiducial one, employs $r_{\rm max}=v\,t$ and $r_{\rm min}=G\, m_\bullet/v^2$, yielding $\ln\Lambda=\ln [v^3(t)\, t/G\,m_\bullet(t)]$; we also check that this prescription brings about very similar results to that based on $r_{\rm max}=2\, r$ and $r_{\rm min}=G\, m_\bullet/v^2$, corresponding to $\ln\Lambda=\ln [v^2(t)\, r(t)/G\,m_\bullet(t)]$

\subsection{Mass accretion onto perturbers}\label{accr}

While the compact remnant, aka the perturber, is moving through the sea of gaseous particle, it can increase its mass by accretion (e.g., Bondi \& Hoyle 1944; Edgar 2004; Cant\'o et al. 2013; Sanchez-Salcedo \& Chametla 2018). Note that in our context the perturber is a compact remnant in supersonic motion and the gain in mass by accretion is expected to be slow, so that we can safely neglect tidal debris effects on the orbit evolution.

Mass accretion causes a net deceleration of the compact remnant
\begin{equation}
a_{\rm acc}=-\frac{\dot{m_\bullet}\,v(t)}{m_\bullet(t)}
\end{equation}
and a simultaneous increase of the dynamical friction force $F_{\rm DF}$, which is proportional to the time-dependent mass $m_\bullet^2(t)$ after Eq.~(\ref{force}). In order to compute the mass accretion rate for a compact object moving through a gaseous medium, we use the recipe by Lee \& Stahler (2011, 2014; see also Tagawa et al. 2016)
\begin{equation}
\frac{{\rm d} m_\bullet}{{\rm d} t}=4\pi G^2\,m_\bullet^2\,\frac{\rho}{c_s^3}\,\frac{\sqrt{\lambda^2+\mathcal{M}^2}}{(1+\mathcal{M}^2)^2}~,
\end{equation}
where $\lambda=1.12$. Since in our context the motion is largely supersonic, including this mass accretion is of minor relevance for what concerns the estimate of the dynamical friction timescales.

\subsection{Orbital decay by gaseous dynamical friction}\label{timescale}

We now compute an estimate of the dynamical friction timescale for a stellar compact remnant to migrate from its initial position toward the galaxy center. The total velocity $v=\sqrt{v_r^2+v_\theta^2}$, the tangential component $v_\theta$ and the distance $r$ from the galaxy center determine the energy and angular momentum per unit mass as
\begin{equation}
\left\{
\begin{aligned}
\varepsilon & =\frac{v^2}{2}+\phi(r)\\
\\
j & =r\,v_\theta~,
\end{aligned}
\right.
\end{equation}
which actually are the basic quantities to follow the orbital evolution. The dynamical friction force $|F_{\rm DF}|$ will dissipate both energy and angular momentum according to the evolution equations
\begin{equation}
\left\{
\begin{aligned}
\frac{{\rm d} \varepsilon}{{\rm d} t} & =-v\frac{|F_{\rm DF}|}{m_\bullet}\\
\\
\frac{{\rm d} j}{{\rm d} t} & =-\frac{j}{v}\frac{|F_{\rm DF}|}{m_\bullet}~.
\end{aligned}
\right.
\end{equation}

\begin{deluxetable*}{llccccccccccccccc}
\tabletypesize{\scriptsize}\tablewidth{0pt}\tablecaption{Gaseous dynamical friction timescale: parameter dependence}
\tablehead{\colhead{Profile} & \colhead{$n$} & \colhead{$\alpha$} & & \colhead{$\ln\Lambda$} & &
\colhead{$\mathcal{N}/10^8$ yr} & \colhead{$a$} & \colhead{$b$} &
\colhead{$c$} & \colhead{$\beta$} & \colhead{$\gamma$} &}\startdata
{\rm Sersic} & $1.5$ & $0.6$ & &  $\ln[v^3(t) t/G m_\bullet(t)]$ & & $3.4$ & $-0.95$ & $0.45$& $-1.2$ & $1.5$ & $2.5$ &\\
\\
{\rm Sersic} & $1.5$ & $1$ & & $\ln[v^3(t) t/G m_\bullet(t)]$ & & $5.9$ & $-0.95$ & $0.45$& $-1$ & $1.5$ & $2.4$ &\\
{\rm Sersic} & $4$ & $0.6$ & & $\ln[v^3(t) t/G m_\bullet(t)]$ & & $13.6$ & $-0.95$ & $0.45$& $-1.2$ & $1.5$ & $2.4$ &\\
\\
{\rm Sersic} & $1.5$ & $0.6$ & & $\ln[R_{e}v^2(t)/G m_\bullet(t)]$ & & $2.5$ & $-0.95$ & $0.45$& $-1.2$ & $1.8$ & $2.6$ \\
\\
{\rm Sersic} & $1.5$ & $0.6$ & & $\ln[R_{e}v^2/G m_\bullet]={\rm const}$ & & $2.2$ & $-1$ & $0.5$& $-1.2$ & $2$ & $2.7$ &\\
{\rm Hernquist} & $-$ & $1$ & & $\ln[R_{e}v^2/G m_\bullet]={\rm const}$ & & $5.7$ & $-1$ & $0.5$& $-1$ & $2$ & $2.5$ &\\
{\rm SIS} & $-$ & $2$ & & $\ln[R_{e}v^2/G m_\bullet]={\rm const}$ & & $21.4$ & $-1$ & $0.5$ & $-0.5$ & $2$ & $2$ &\\
\enddata
\label{tabella1}
\end{deluxetable*}

We use the orbit-averaged approximations (e.g., Lacey \& Cole 1993; Tonini et al. 2006), yielding
\begin{equation}
\left\{
\begin{aligned}
\langle\dot\varepsilon\rangle & =-\frac{\int_{r_-}^{r_+}{\rm d} r\, (v/v_r)\, |F_{\rm DF}|/m_\bullet}{\int_{r_-}^{r_+}{\rm d} r/v_r}\\
\\
\langle\dot j\rangle & =-j\frac{\int_{r_-}^{r_+}{\rm d} r\, (1/v_r)\, (1/v)\, |F_{\rm DF}|/m_\bullet}{\int_{r_-}^{r_+}{\rm d} r/v_r}~,
\end{aligned}
\right.
\end{equation}
where $v_r=\sqrt{2\,[\varepsilon-\phi(r)]-j^2/r^2}$ is the radial velocity component, and $r_-$ and $r_+$ are the pericenter and apocenter radial positions determined by the condition $v_r=0$; the corresponding orbital eccentricity can be computed as
\begin{equation}
e=\frac{r_+-r_-}{r_+ + r_-}~.
\end{equation}

Note that when in the Coulomb logarithm $\ln\Lambda=\ln r_{\rm max}/r_{\rm min}$ a time-dependent $r_{\rm max}(t)=v\, t$ is adopted, the above equation must be modified somewhat. We recall that this choice of $r_{\rm max}$ was justified by Ostriker (1999) as the displacement of a perturber travelling on a straight line after a time $t$, so it represents a lengthscale of the wake. In the case of elliptical orbits, such a quantity depends on the perturber position and, since at the apocenter and pericenter the direction of motion is reversed, the wake cannot be longer than half of the orbit. Thus we divide the above orbit-averaged integral into two halves, taking into account that when the perturber is at apocenter or pericenter the time appearing into the expression for $r_{\rm max}\propto t$ must be reset to zero. Writing $F_{\rm DF}(\ln\Lambda)$ as a function of the Coulomb logarithm, we use
\begin{widetext}
\begin{equation}
\left\{
\begin{aligned}
\langle\dot\varepsilon\rangle &=-\frac{\int_{r_-}^{r_+}{\rm d} r(v/v_r)|F_{\rm DF}(\ln\Lambda_-)|+\int_{r_+}^{r_-}{\rm d} r(v/v_r)|F_{\rm DF}(\ln\Lambda_+)|}{2m_\bullet\int_{r_-}^{r_+}{\rm d} r/v_r}\\
\\
\langle\dot j\rangle &=-j\frac{\int_{r_-}^{r_+}{\rm d} r(1/v_rv)|F_{\rm DF}(\ln\Lambda_-)|+\int_{r_+}^{r_-}{\rm d} r(1/v_rv)|F_{\rm DF}(\ln\Lambda_+)|}{2m_\bullet\int_{r_-}^{r_+}{\rm d} rv_r}~,
\end{aligned}
\right.
\end{equation}
\end{widetext}
where $\ln \Lambda_{\pm}=\ln (v\,t_{\pm}/r_{\rm min})$ in terms of the time $t_{\pm}(r)=\int^{r}_{r_{\pm}}{\rm d} r/v_r$ elapsed at distance $r$ from/to pericenter/apocenter.

\begin{deluxetable*}{ccccccccccccccccc}
\tabletypesize{\scriptsize}\tablewidth{0pt}\tablecaption{Gaseous dynamical friction timescale: examples}
\tablehead{\colhead{$r/$pc} & \colhead{$v_{r}/\sigma(r)$, $v_\theta/\sigma(r)$} & & \colhead{$r_c/$pc} & \colhead{$j/j_c$} & & &\multicolumn{4}{c}{$\tau_{\rm DF}/$Gyr}\\
\\
\cline{8-12}
& & & & & & &\colhead{$m_\bullet=1.5\, M_\odot$} & \colhead{$m_\bullet=10\, M_\odot$} & \colhead{$m_\bullet=40\, M_\odot$} & \colhead{$m_\bullet=100\, M_\odot$}}
\startdata
5 & 1, 1 & & 30 & 0.18 & & & $-$ & 4.1 & 1.1 & 0.46 &\\
5 & 1, 0.1 & & 18 & 0.041 & & & 0.75 & 0.12 & 0.033 & 0.014 & \\
5 & 0.1, 1 & & 18 & 0.41 & & & $-$ & 4.0 & 1.0 & 0.44 &\\
5 & 0.1, 0.1 & & 4 & 0.55 & & & 0.75 & 0.12 & 0.033 & 0.014 &\\
\\
15 & 1, 1 & & 50 & 0.31 & & & $-$ & $-$ & $9.4$ & $3.9$ & \\
15 & 1, 0.1 & & 33 & 0.064 & & & $6.4$ & $1.0$ & $0.28$ & $0.12$ &\\
15 & 1, 0.025 & & 33 & 0.016 & & & $0.80$ & $0.13$ & $0.035$ & $0.015$ &\\
\\
30 & 1, 1 & &76 & 0.41 & & & $-$ & $-$ & $-$ & $-$ & \\
30 & 1, 0.1 & & 50 & 0.080 & & & $-$ & $4.1$ & $1.1$ & $0.46$ &\\
30 & 1., 0.01 & &50 &0.0081 & & & $0.79$ & $0.13$ & $0.035$ & $0.015$ &\\
\\
50 & 1, 1 & &100 & 0.49 & & & $-$ & $-$ & $-$ & $-$ &\\
50 & 1, 0.1 & &70 & 0.089 & & & $-$ & $-$ & $3.0$ & $1.2$ &\\
50 & 1, 0.01 & &70 & 0.0090 & & & $2.1$ & $0.35$ & $0.093$ & $0.039$ &\\
50 & 1, 0.005 & &70 & 0.0045 & & & $0.75$ & $0.12$ & $0.033$ & $0.014$ &\\
\\
150 & 1, 1 & &200 & 0.64 & & & $-$ & $-$ & $-$ & $-$ &\\
150 & 1, 0.1 & &150 & 0.10 & & & $-$ & $-$ & $-$ & $10$ &\\
150 & 1, 0.01 & &150 & 0.010 & & & $-$ & $2.8$ & $0.75$ & $0.31$ &\\
150 & 1, 0.001 & &150 & 0.0010 & & & $0.55$ & $0.091$ & $0.024$ & $0.010$ &\\
\\
300 & 1, 0.1 & &26 & 0.10 & & & $-$ & $-$ & $-$ & $-$ &\\
300 & 1, 0.01 & &26 & 0.010 & & & $-$ & $-$ & $3.0$ & $1.2$ &\\
300 & 1, 0.001 & &26 & 0.0010 & & & $2.0$ & $0.33$ & $0.088$ & $0.037$ &\\
300 & 1, 0.0005 & &26 & 0.00050 & & & $0.72$ & $0.12$ & $0.032$ & $0.013$ &\\
\enddata
\tablecomments{A dash ($-$) indicates a timescale longer than $10$ Gyr.}
\label{tabella2}
\end{deluxetable*}

A couple of consequences found from computing the above terms are the following. First, the points which contribute more to the dynamical friction force $F_{\rm DF}\propto \rho/v^2$ turn out to be the pericenter and apocenter; the former is the innermost point of the orbit where the gas density $\rho$ is higher, while the latter is the outermost point of the orbit where the velocity $v$ of the perturber is smaller. Second, gaseous dynamical friction is much more efficient in dissipating angular momentum than energy; as a consequence, the apocenter $r_+$ evolves slowly (being mainly determined by the orbital energy), while the pericenter $r_-$ decays much rapidly (being directly related to the centrifugal barrier) and the overall orbit eccentricity increases. In fact, this process is of runaway type since as $j$ decreases, more higher density regions are reached at pericenter while the velocity lowers near apocenter, to imply enhanced dynamical friction force and further angular momentum loss.

We numerically integrate the orbit-averaged equations ${\rm d} \varepsilon/{\rm d} t = \langle\dot \varepsilon\rangle$ and ${\rm d} j/{\rm d} t = \langle\dot j\rangle$ to determine the timescale $\tau_{\rm DF}$ needed for the compact remnant to migrate to the galaxy center. Actually, we halt the computation when the pericenter attains a value below $r_-\sim 10^{-5}$ pc, since in these nuclear region the migrating compact remnant feels the potential of the growing central BH, and rapid energy and angular momentum losses eventually take place due to emission of GWs; the orbit-averaged loss rates (see Peters 1964)
$\langle\dot \varepsilon_{\rm GW}\rangle\propto (1-e^2)^{-7/2}$ and $\langle \dot j_{\rm GW}\rangle\propto (1-e^2)^{-2}$ are very efficient since the remnant tends to reach such inner regions with high eccentricity $e\approx 1$, enforced by the gaseous dynamical friction on larger scales; subsequently, the orbit shrinks rapidly and merging between the central mass and the stellar remnant can occur. We stress that the runaway nature of the pericenter decay makes of minor relevance the choice of the minimum radius where the computation of the dynamical friction evolution is stopped and $\tau_{\rm DF}$ is evaluated.

In Fig.~\ref{fig:dynfric} we show the evolution of the potential and total energy, of the pericenter and apocenter, and of the orbital eccentricity, for two representative cases with nearly circular and mildly eccentric initial orbits; reference values $M_{\rm gas}=10^{11}\, M_\odot$, $R_e=1$ kpc and $m_\bullet=100\, M_\odot$ have been adopted. The reader can easily appreciate the runaway decrease of the pericenter $r_-$; this is mainly driven by the loss in angular momentum, which reduces the centrifugal barrier and hence
flattens the shape of the effective potential at small radii. Contrariwise, the apocenter $r_+$ is mainly determined by the decrease in total energy and decays slowly; as a consequence, the orbital eccentricity increases with time. Less eccentric initial conditions imply longer overall dynamical timescales, but a more rapid evolution of the apocenter (somewhat parallel to the pericenter), since the system remain quite close to a circular orbit, with the total energy hovering around the minimum of the effective potential; correspondingly, the eccentricity stays low for most of the evolution, and then rises abruptly close to the pericenter runaway.

\subsection{Gaseous dynamical friction timescales}

The resulting dynamical friction timescale $\tau_{\rm DF}$ depends on the properties of the background gas mass distribution (half-mass radius $R_e$, total mass $M_{\rm gas}$ and shape parameters $n$ and $\alpha$), and on the initial mass $m_\bullet$, energy $\varepsilon$ and angular momentum $j$ of the compact remnant. Actually it is convenient to express the dependence on energy through the circular radius $r_c(\varepsilon)$ that the compact remnant would have if it were on a circular orbit at given energy $\varepsilon$; this is computed just by solving $\varepsilon=G\, M(<r_c)/2\, r_c+\phi(r_c)$.
On the same footing $j_c(\varepsilon)=\sqrt{G\, M(<r_c)\, r_c}$ will be the angular momentum associated to that circular orbit, and so the ratio $j/j_c(\varepsilon)$ constitutes a measure of the (non-)circularity of the motion. In terms of these quantities, the dynamical friction timescale can be expressed as
\begin{widetext}
\begin{equation}
\tau_{\rm DF} =\mathcal{N}\,\left(\frac{m_\bullet}{100M_\odot}\right)^{a}\, \left(\frac{M_{\rm gas}}{10^{11}\,M_\odot}\right)^{b}\, \left(\frac{R_e}{1\,\rm kpc}\right)^{c}\,\left[\frac{j}{j_c(\varepsilon)}\right]^{\beta}\, \left[\frac{r_c(\varepsilon)}{10\, \rm pc}\right]^{\gamma}~,
\label{taudynamical}
\end{equation}
\end{widetext}
where $\mathcal{N}$ is a normalization constant. When the mass accretion onto the perturber is neglected and the Coulomb logarithm $\ln\Lambda=\ln[R_e\, v^2/G\, m_\bullet]$ in the dynamical friction force is taken to be constant in time (see Sect.~\ref{dynfriction}), one obtains the exponents $a\approx -1$, $b\approx 1/2$, $c=(\alpha-3)/2$,  $\beta\approx 2$, and $\gamma\approx 3-\alpha/2$. The dependencies on $m_\bullet$ and $M_{\rm gas}$ are somewhat trivial and can be derived basing on a simple dimensional analysis of the orbital equations (e.g., Lacey \& Cole 1993; Tamburello et al. 2017). In addition, the dependencies on $R_e$ and on $r_c(\varepsilon)$ are controlled by the inner slope of the density profile $\alpha$, independently on the Mach number (and actually being the same also for dynamical friction against a collisionless background); for example, adopting a SIS profile with $\alpha=2$ as in Lacey \& Cole (1993) yields $c\approx -1/2$ and $\gamma\approx 2$, in agreement with their result, while adopting a Hernquist profile with $\alpha=1$ yields $c\approx -1$ and $\gamma\approx 5/2$ as in Tamburello et al. (2017). Finally, the exponent $\beta$ is found to be independent of the profile, but to depend crucially on the Mach number of the perturber during its motion in the background gaseous atmosphere; in particular, if the motion is supersonic like in our case then $\beta\approx 2$ applies, while for (sub)sonic motion (or when the medium is collisionless) the dependence is found to be much shallower $\beta\approx 0.78$, as in Lacey \& Cole (1993).

When the mass accretion onto the perturber is switched on and a time-dependent Coulomb logarithm $\ln\Lambda=\ln[R_e\, v^2(t)/G\, m_\bullet(t)]$ is considered, the exponents in Eq.~(\ref{taudynamical}) changes into $a\approx -0.95$, $b\approx 0.45$, $c\approx (\alpha-3)/2$, $\gamma\approx 3-\alpha/2+\epsilon_0(n,\alpha)$, and $\beta\approx 1.8$; here $\epsilon_0(n,\alpha)$ is a small correction dependent on the shape parameter of the density profile, with typical values $\epsilon_0(n,\alpha)\sim 10^{-1}$. Finally, when our fiducial expression of the Coulomb logarithm $\ln\Lambda=\ln[v^3(t)\, t/G\, m_\bullet(t)]$ is implemented, one finds the exponents $a\approx -0.95$, $b\approx 0.45$, $c\approx (\alpha-3)/2$, $\gamma\approx 3-\alpha/2+\epsilon_1(n,\alpha)$, and $\beta\approx 1.5+\epsilon_2(j/j_c)$. Thus in this case not only $\gamma$ gets a correction $\epsilon_1(n,\alpha)\sim 10^{-1}$ dependent on the shape of the density profile, but also $\beta$ acquires a weak dependence on the circularity $j/j_c(\varepsilon)$ via the quantity $\epsilon_2(j/j_c)$ that spans the range $-0.3$ to $0$ to $0.3$ when $j/j_c$ increases from $0$ to $0.5$ to $1$. In Table~\ref{tabella1} we report the values of the exponents and of the normalization constant $\mathcal{N}$ appearing in Eq.~(\ref{taudynamical}) for some representative cases.

In Table~\ref{tabella2} we present specific examples of the resulting dynamical friction timescales. The Table refers to the reference Sersic density profile with $n=1.5$, $\alpha=0.6$, and time-dependent Coulomb logarithm $\ln[v^3(t) t/G m_\bullet(t)]$ with mass accretion onto the perturber switched on. For different values of the initial physical radius $r$, velocities $v_{r,\theta}/\sigma(r)$ and perturber mass $m_\bullet$, we report the circular radius $r_c(\varepsilon)$, circularity $j/j_c(\varepsilon)$ and the dynamical friction timescales $\tau_{\rm DF}$. We find that dynamical friction timescales smaller than $1$ Gyr are allowed for a variety of initial conditions and remnant masses, implying that the process can be relevant for the formation of heavy BH seeds.

The dependence of the dynamical friction timescale on initial conditions is easily explained. At given initial radius, raising $v_r$ increases the energy so enhancing $r_c$ but at the same time it decreases the circularity, so that the overall dependence on $v_r$ is weak; this is why in the Table $v_r$ is changed only in the case referring to $r=5$ pc, but the behavior for other radii stays put. The impact of $v_\theta$ is significant, since decreasing it both reduces the energy and the circularity, so shortening $\tau_{\rm DF}$. Increasing the initial radius $r$ basically enhances the energy so raising $r_c(\varepsilon)$, making $\tau_{\rm DF}$ longer.

In Table~\ref{tabella2} we also highlight that at larger radii the dynamical friction timescale can still be appreciably smaller than $1$ Gyr if $v_\theta$ is sufficiently small. Given the Gaussian shape of the tangential velocity distribution (see Eq.~\ref{dpdv}), this implies that a lower fraction of the compact remnants produced at larger radii can reach the nuclear region and contribute to the growth of the central BH seed. On the other hand, given the inner power-law shape of the gas density profile, the number of compact remnants produced in larger radial shells increases (see Eq.~\ref{dpdr}). All in all, we find that these two effects partially compensate, so as to cause a similar overall contribution to the central BH mass growth from remnants formed at different radii, at least out to $r\gtrsim 300$ pc where the exponential cutoff of the density profile progressively takes over reducing drastically the number of available remnants.

\section{Merging rates and central mass growth}\label{central mass}

We now exploit the expression for the dynamical friction timescale derived in the previous Section to compute the merging rate of compact remnants at different galactic ages, so evaluating the contribution of this process to the growth of the central supermassive BH seed; in the next section we will discuss how the efficiency of BH growth by dynamical friction compares and couples with that due to standard disk (Eddington-like) accretion.

The merging rate per unit compact remnant mass due to dynamical friction at a galactic age $\tau$ inside a galaxy with spatially-integrated SFR $\psi$ at redshift $z$ can be written as
\begin{equation}
\begin{aligned}
\frac{{\rm d}\dot N_{\rm DF}}{{\rm d}m_\bullet}&(m_\bullet,\tau|\psi,z)=\int{\rm d} r\frac{{\rm d} p}{{\rm d} r}(r)\int{\rm d} v_\theta\frac{{\rm d} p}{{\rm d} v_\theta}(v_\theta|r)\times\\
\\
&\times\int{\rm d} v_r\frac{{\rm d} p}{{\rm d} v_r}(v_r|r)\,R_{\rm birth}(m_\bullet, \tau-\tau_{\rm DF}|\psi,z)~;
\end{aligned}
\label{rmergerv}
\end{equation}
here ${\rm d} p/{\rm d} r$ and ${\rm d} p/{\rm d} v_{r,\theta}$ are the probability distributions of initial radii and velocities given in Eqs. (\ref{dpdr}) and (\ref{dpdv}); $R_{\rm birth}(m_\bullet, \tau|\psi,z)$ is the birthrate for a stellar compact remnant of mass $m_\bullet$ at a galactic age $\tau$ and $\tau_{\rm DF}[m_\bullet, \varepsilon(r,v_\theta,v_r), j(r,v_\theta)]$ is the dynamical friction timescale (see Eq. \ref{taudynamical}) for a compact remnant of mass $m_\bullet$, formed at radius $r$ with initial velocities $v_{r,\theta}$, or equivalently with energy $\varepsilon$ and angular momentum $j$. The underlying rationale of this expression is that the merging rate at the galactic age $\tau$ depends directly on the birthrate $R_{\rm birth}$ at a galactic age $\tau-\tau_{\rm DF}$ (plainly we require that $\tau_{\rm DF}$ for a compact remnant is longer than the progenitor star's lifetime); the resulting quantity turns out to be a function of the initial radius and velocities, that are averaged over the associated distributions.

Notice that in the above we have parameterized the merging rates in terms of the spatially-integrated SFR of an ETG progenitor, since in the next Sections this will ease the computation of cosmic average quantities via SFR-based galaxy statistics. As already shown, both the dynamical friction timescale $\tau_{\rm DF}$ and the distributions of initial radii and velocities depend on the gas density profile, and in particular on the initial total gas mass $M_{\rm gas}$ (see Eqs.~\ref{taudynamical}-\ref{dpdr}-\ref{dpdv}). We compute $M_{\rm gas}$  for a given value of $\psi$ by first estimating the stellar mass from the redshift-dependent galaxy main sequence relation $\psi-M_\star$ by Speagle et al. (2014), and then inferring the initial gas mass from the redshift-dependent $M_\star/M_{\rm gas}-M_\star$ relation by Lapi et al. (2017; see also Moster et al. 2013; Aversa et al. 2015; Shi et al. 2017; Behroozi et al. 2019) based on abundance matching techniques.

\begin{figure*}
\centering
\includegraphics[width=.49\textwidth]{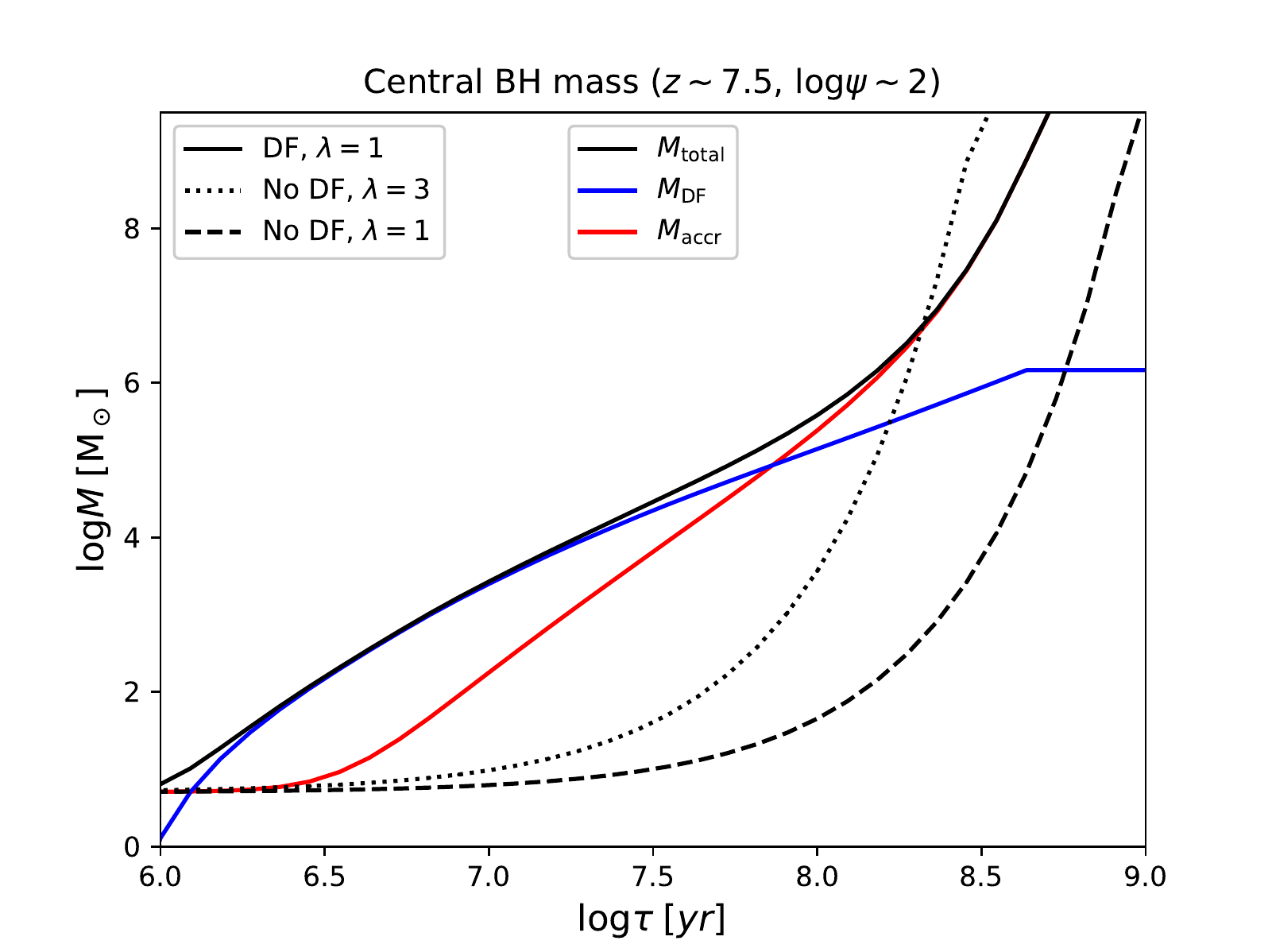}
\includegraphics[width=.49\textwidth]{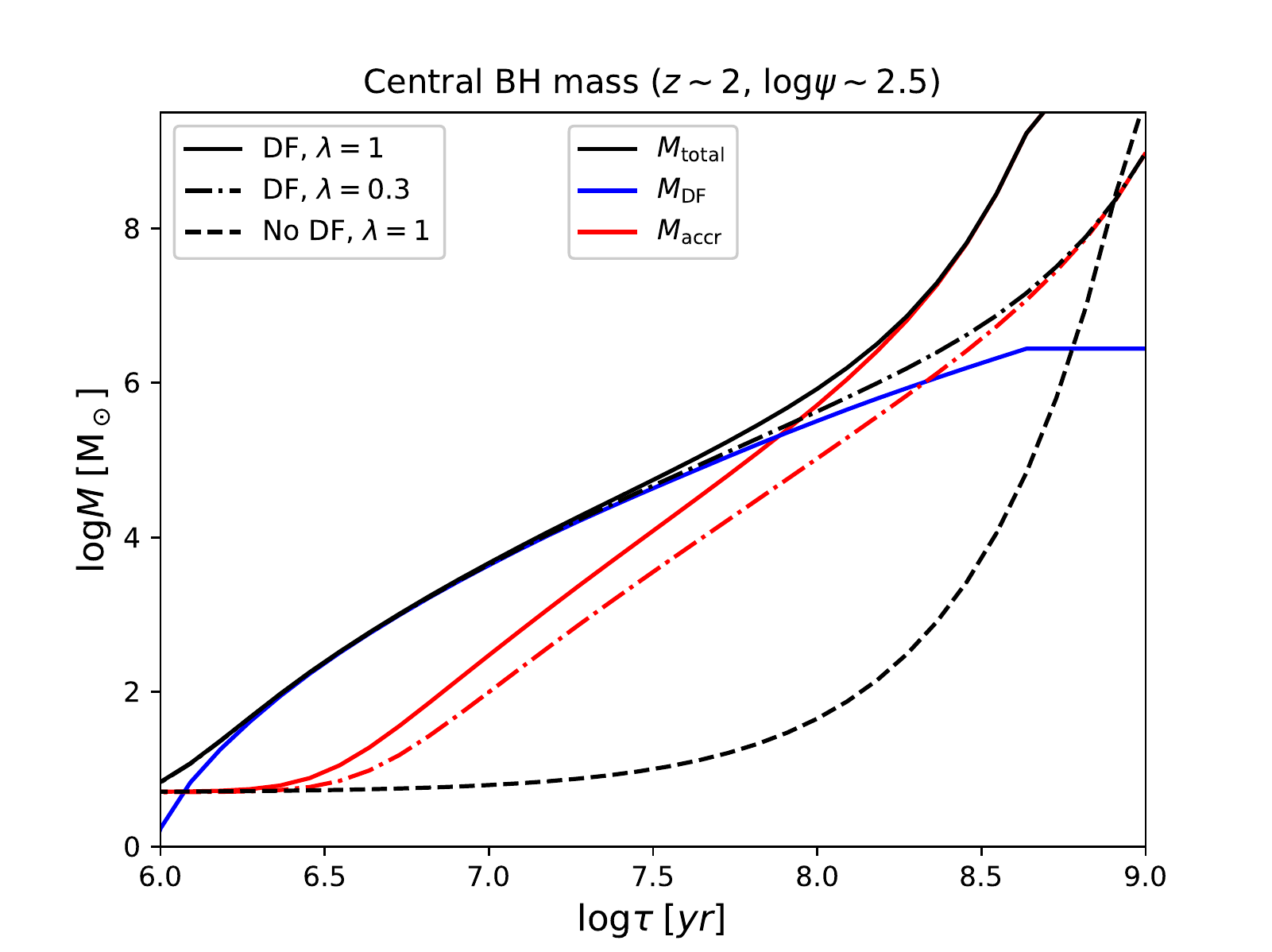}
\caption{Left panel: evolution of the central BH mass (black solid lines) for a galaxy located at $z\sim 7.5$ and featuring a SFR $\psi\sim 100\, M_\odot$ yr$^{-1}$, representative of the typical host of the most distant quasars. The contribution to the hole growth from gaseous dynamical friction (blue solid lines) and from disk accretion with Eddington ratio $\lambda=1$ (red solid lines) is highlighted; for comparison, evolutionary tracks due to pure disk accretion with $\lambda=1$ (dashed black line) and $\lambda=3$ (dotted black line) are also illustrated. Right panel: the same is shown for a typical ETG progenitor located at $z\sim 2$ and featuring SFR $\psi\sim 300\, M_\odot$ yr$^{-1}$; we illustrate the evolution when including dynamical friction and disk accretion with $\lambda=1$ (solid lines) or $\lambda=0.3$ (dot-dashed lines), and that for pure disk accretion with $\lambda=1$ (dashed lines).}
\label{fig:mass}
\end{figure*}

Coming back to Eq.~(\ref{rmergerv}) the birthrate $R_{\rm birth}$ is computed as follows (e.g., Dvorkin et al. 2016; Cao et al. 2018; Li et al. 2018; Boco et al. 2019):
\begin{equation}
\begin{aligned}
R_{\rm birth}&(m_\bullet,\tau|\psi,z)=\psi\,\int_{m_{\star,\rm min}}{\rm d} m_\star\,\phi(m_\star)\times\\
\\
&\times\frac{{\rm d} p}{{\rm d} m_\bullet}[m_\bullet|m_\star,Z(\tau|\psi,z)]~;
\end{aligned}
\label{rbirth}
\end{equation}
the quantity ${\rm d} p/{\rm d} m_\bullet$ represents the probability distribution of producing a compact remnant of mass $m_\bullet$ given the initial star mass $m_\star$ and a metallicity $Z$. Following Boco et al. (2019) this probability distribution is taken to be a lognormal
\begin{equation}
\begin{aligned}
\frac{{\rm d}p}{{\rm d}\log m_\bullet}&(m_\bullet|m_\star,Z) = \frac{1}{\sqrt{2\pi}\,\sigma_{\log m_\bullet}}\times\\
\\
&\times\exp\left\{-[\log m_\bullet -\log m_\bullet(m_\star,Z)]^2/2\,\sigma_{\log m_\bullet}^2\right\}~.
\end{aligned}
\label{eq|remnant}
\end{equation}
centered around the average relationship $m_\bullet(m_\star,Z)$ obtained by Spera \& Mapelli (2017; see also Spera et al. 2015 for details) via the \texttt{SEVN} stellar evolutionary code
including pair-instability and pair-instability pulsational supernovae (causing a `failed' explosion and a direct collapse to BH), and with a dispersion of $\sigma_{\log m_\bullet}=0.1$ dex that takes into account plausible astrophysical uncertainties and intrinsic scatter. The Spera relation $m_\bullet(m_\star,Z)$, relating the mass of the compact remnant to that of the progenitor star, depends crucially on the gas metallicity $Z(\tau|\psi,z)$; we compute the latter as a function of the galactic age $\tau$ using the detailed chemical galaxy evolution models by Pantoni et al. (2019). The outcome is a rapid (almost linear) increase of the metallicity with galactic age up to a saturation value dependent on the galaxy SFR $\psi$ and redshift $z$; such a model has been shown to reproduce both the metal enrichment properties of ETGs and their high-$z$ star-forming progenitors. In Eq.~(\ref{rbirth}) the remnant distribution ${\rm d} p/{\rm d} m_\bullet$ is then integrated over the star masses $m_\star$, weighting by the IMF $\phi(m_\star)$ from the lower limit $m_{\star,\rm min}\sim 8\, M_\odot$ required to produce a compact remnant (i.e., neutron star or BH). Finally, the result is multiplied by the SFR $\psi$ which just specifies that galaxies with larger SFRs will produce more numerous compact remnants.

In Fig.~\ref{fig:mergerrate} we illustrate the merger rate ${\rm d}\dot N_{\rm DF}/{\rm d}\log m_\bullet$ per unit logarithmic bin of compact remnant mass $m_\bullet$, for a galaxy with redshift $z=2$ and spatially-integrated SFR $\psi=300\,M_\odot/\rm yr$, at different galactic ages $\tau$; this SFR is a typical value for a star-forming ETG progenitors at $z\sim 2$, that characterizes galaxies at the knee of the SFR function (e.g., Gruppioni et al. 2013; 2015; Mancuso et al. 2016; Lapi et al. 2017, 2018). At early times (say $10^7$ yr, which are anyway needed for the most massive stars to explode as supernovae) only the most massive compact remnants with $m_\bullet\gtrsim 30\, M_\odot$ contribute to the merging rate, since the dynamical friction timescale is shorter for them (see Eq.~\ref{taudynamical} and Table~\ref{tabella2}). At later stages, compact remnants of all masses progressively enter into the game. After some $10^7$ yr the shape of the merging rate becomes stationary, with some relevant characteristic features: (i) a peak at around $m_\bullet\sim 1.5-2\,M_\odot$ representing the contribution from neutron stars, which are much more abundant than BH for the standard Chabrier IMF adopted here; (ii) a rise toward more massive remnants due to the increased efficiency of the dynamical friction process for larger $m_\bullet$; (iii) a second peak for masses in the range $m_\bullet\sim 40-60\,M_\odot$, which are created more frequently according to the birthrate mass spectrum; (iv) a subsequent decline for remnants with $m_\bullet\gtrsim 60\, M_\odot$, that is due to the strong suppression in the birthrate for these masses by pair-instability and pair-instability pulsational supernovae.

As the galaxy age increases, the overall merger rate grows in normalization just because even compact remnants with larger $\tau_{\rm DF}$ can reach the galaxy center. At galactic ages $\tau\gtrsim 10^8$ yr the aforementioned second peak tends to shift toward lower masses, and the drop kicks in for masses $m_\bullet\geq 40\, M_\odot$; this occurs because the metallicity increases with the galactic age, up to a value $Z\gtrsim 0.1\, Z_\odot$ when very massive BH remnants are no more efficiently produced according to the relation $m_\bullet(m_\star,Z)$.

\begin{figure*}
    \centering
    \includegraphics[width=.49\textwidth]{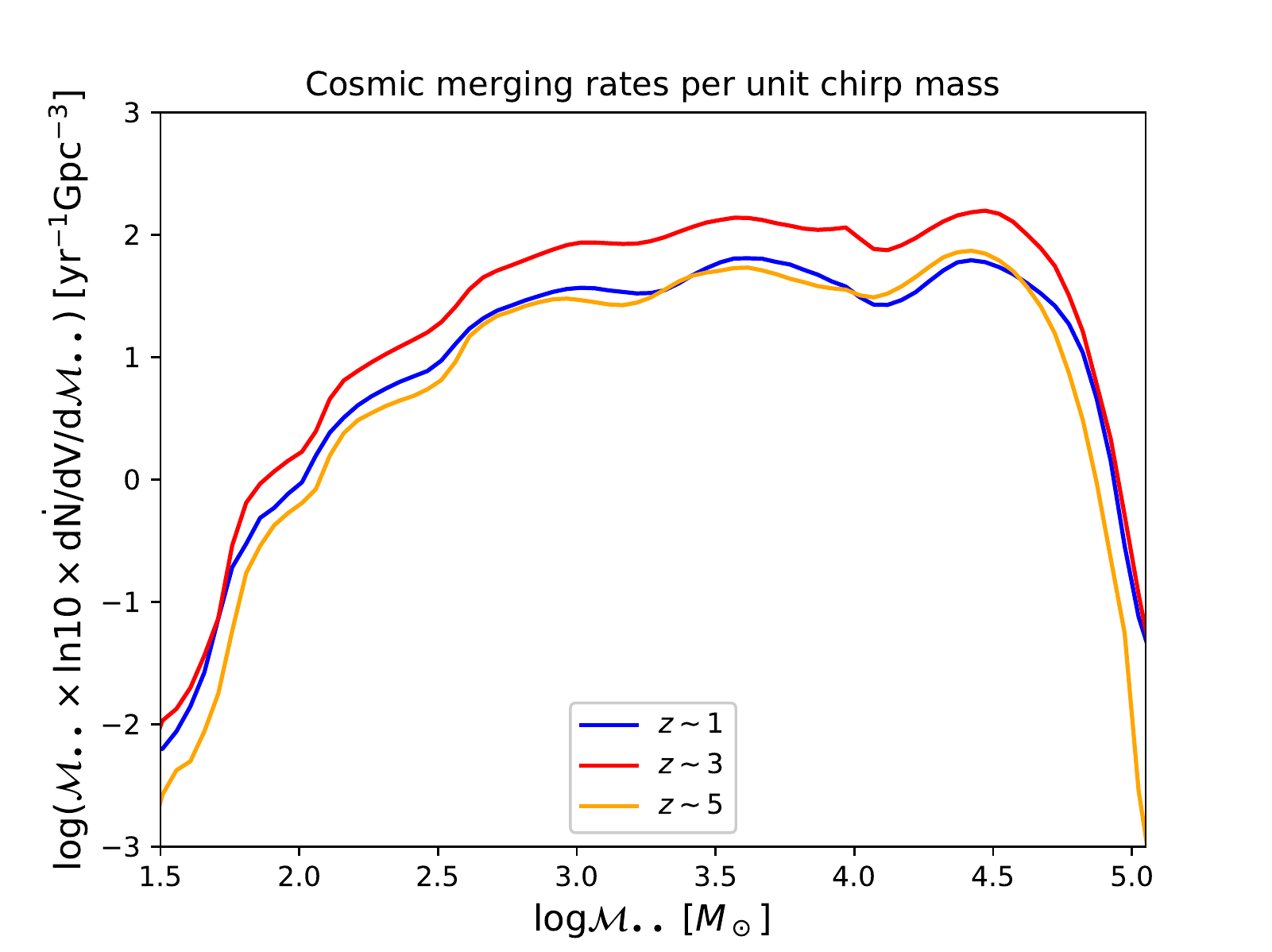}
    \includegraphics[width=.49\textwidth]{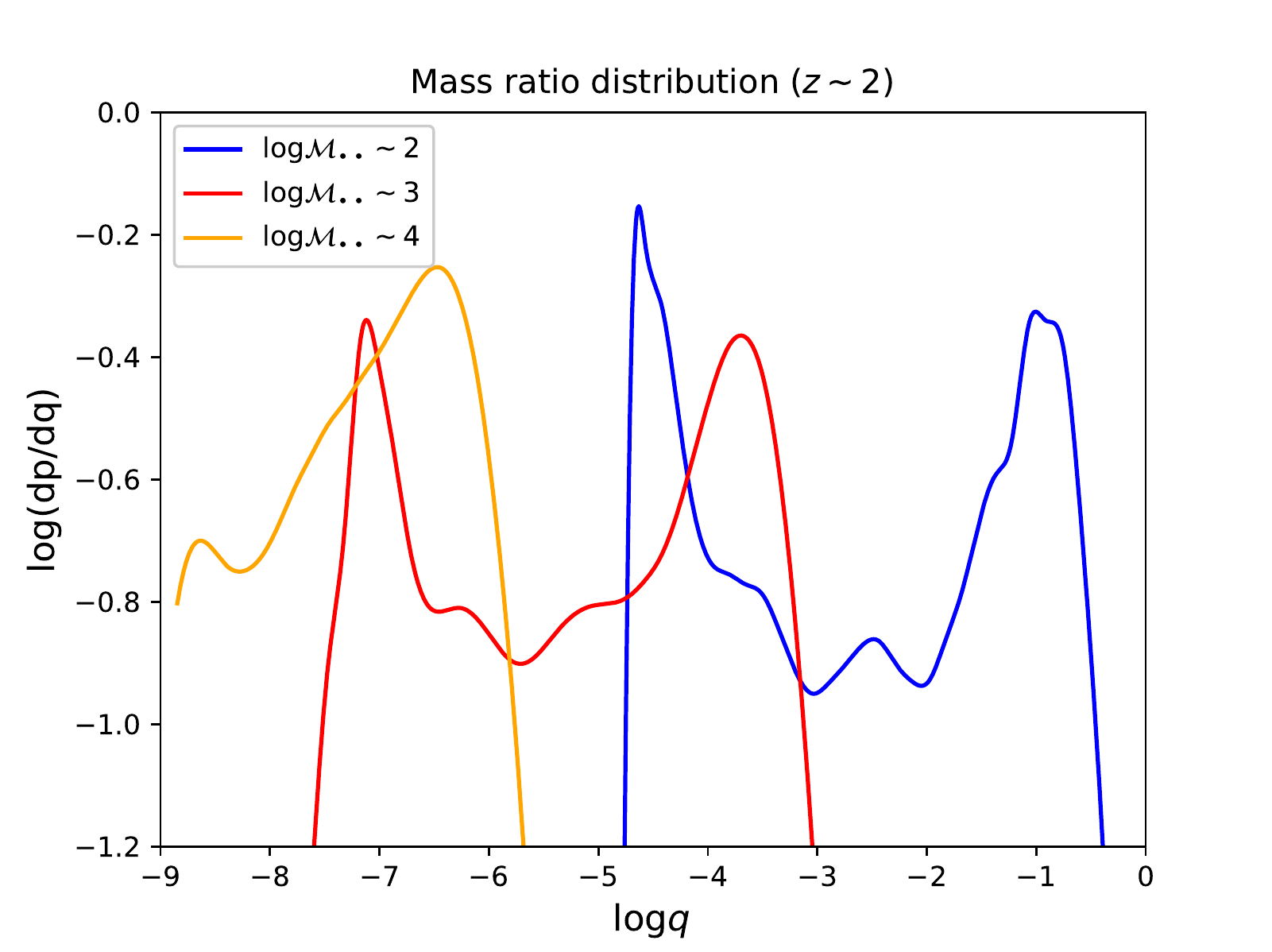}
    \caption{Left panel: cosmic rates ${\rm}d\dot N_{\rm DF}/{\rm d}V\,{\rm d}\log \mathcal{M}_{\bullet\bullet}$ per unit chirp mass of the merging events due to gaseous dynamical friction at redshift $z\sim 1$ (blue line), $3$ (red) and $5$ (orange). Right panel: mass ratio distribution ${\rm d}p/{\rm d}q (q|\mathcal{M}_{\bullet\bullet})$ at redshift $z\sim 2$ for different chirp masses $\mathcal{M}_{\bullet\bullet}\sim 10^2\, M_\odot$ (blue line), $10^3\, M_\odot$ (red), and $10^{4}\, M_\odot$ (orange).}
    \label{fig:chirpq}
\end{figure*}

\subsection{Central BH growth via dynamical friction and disk accretion}\label{accretion}

The overall merging rate at galactic age $\tau$ due to dynamical friction migration of stellar compact remnants can be found by integrating Eq.~(\ref{rmergerv}) over the remnant masses
\begin{equation}
\dot N_{\rm DF}(\tau)=\int{\rm d} m_\bullet\, \frac{{\rm d}\dot N_{\rm DF}}{{\rm d}m_\bullet}(m_\bullet,\tau|\psi,z)
\label{rmergetau}
\end{equation}
while the growth rate of the central BH mass is given by
\begin{equation}
\dot{M}_{\bullet, \rm DF}(\tau)=\int{\rm d} m_\bullet\, m_\bullet\, \frac{{\rm d}\dot N_{\rm DF}}{{\rm d}m_\bullet}(m_\bullet,\tau|\psi,z)~.
\label{ratemass}
\end{equation}
Clearly, further integrating the latter equation over time provides the contribution of dynamical friction to the growth of the central BH as a function of galactic age
\begin{equation}
M_{\bullet,\rm DF}(\tau)=\int_0^\tau{\rm d}\tau'\, \dot{M}_{\bullet, \rm DF}(\tau'|\psi,z)~;
\label{masstau}
\end{equation}
to ease the notation we have dropped from the quantities on the left hand side the explicit dependence on the galaxy SFR $\psi$ and redshift $z$, but the reader should keep track of that for use in the next sections. Note that, for the sake of simplicity, we are assuming that the migrating remnants accumulate their mass in a single (or at least dominant) central BH; actually, in the early stages multiple of such sinks could originate but dynamical friction (being stronger for more massive perturbers) should enforce rapid merging among them.

\begin{figure*}
    \centering
    \includegraphics[width=0.65\textwidth]{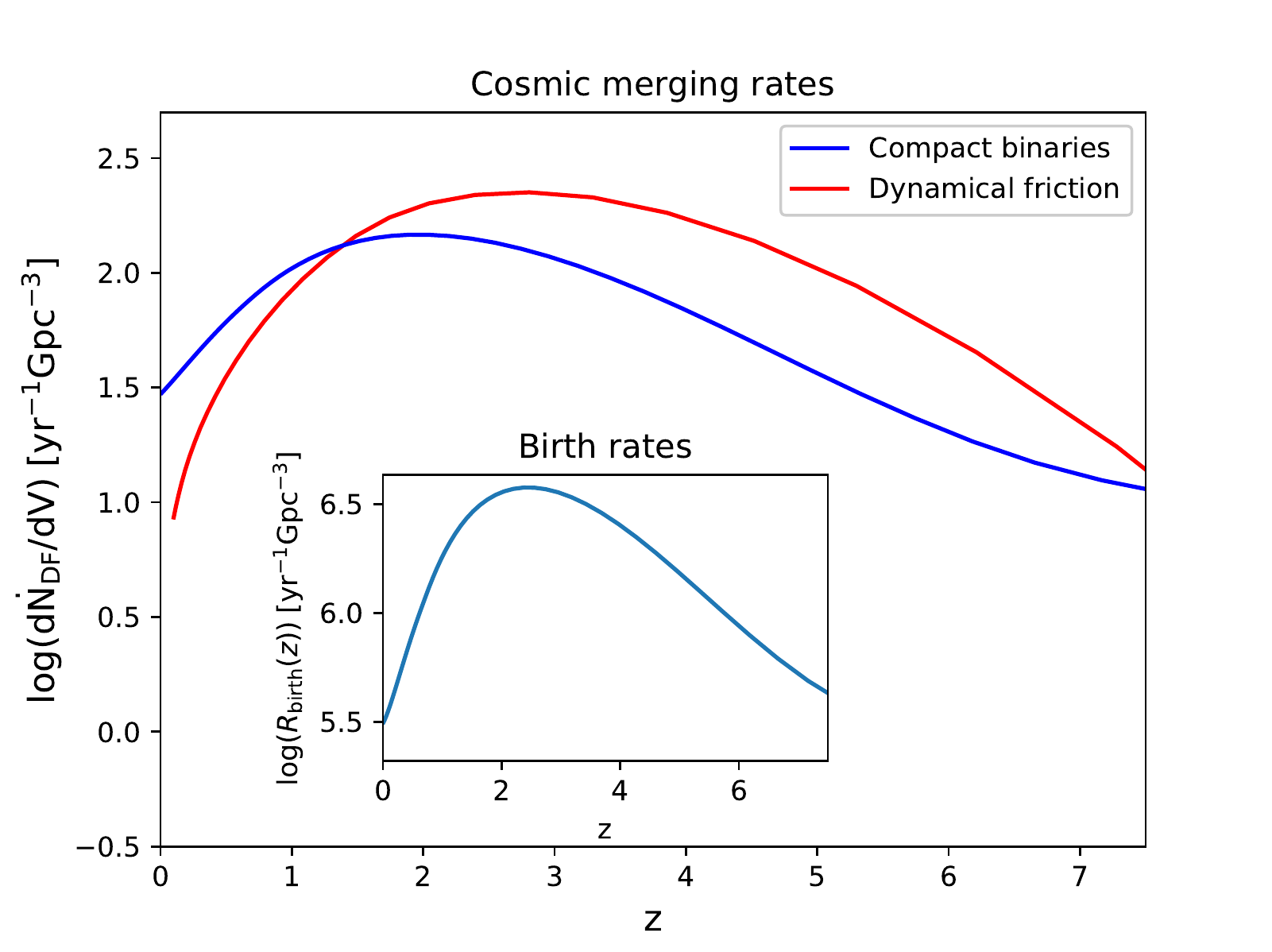}
    \caption{Cosmic merging rate as a function of redshift due to dynamical friction (red line), compared with the analogous quantity for the merging of compact binaries (blue line) in galaxies; the inset shows the underlying cosmic birthrate of stellar compact remnants in galaxies (cyan line).}
    \label{fig:confronto_mergerrate}
\end{figure*}

Once the central BH mass starts to accumulate, standard disk accretion becomes an additional source for the hole growth. In a gaseous-rich environment like the nuclear region of ETG progenitors, the disk accretion is typically demand-limited. For the sake of definiteness, we assume an Eddington-like accretion rate (i.e., proportional to the BH mass) with a given Eddington ratio $\lambda\equiv L/L_{\rm Edd}$ in terms of the Eddington luminosity $L_{\rm Edd}\approx 1.4\times 10^{38}\, M_\bullet/M_\odot$ erg s$^{-1}$, and a radiative efficiency $\eta\equiv L/\dot M_\bullet\, c^2$ of order $10\%$ (see Davis \& Laor 2011; Raimundo et al. 2012; Wu et al. 2013; Aversa et al. 2015). The resulting growth rate due to accretion onto the disk is
\begin{equation}
\dot{M}_{\rm \bullet,\rm acc} = M_\bullet/\tau_{\rm ef}
\end{equation}
where the $e-$folding time amounts to $\tau_{\rm ef}\approx 4.5\times 10^7\, \lambda^{-1}$ yr.

Thus the central mass growth, including both dynamical friction and disk accretion, can be computed simply by integrating the linear differential equation $\dot M_\bullet=\dot{M}_{\bullet, \rm DF}(\tau)+\dot{M}_{\bullet,\rm acc}[M_\bullet(\tau)]$, which yields
\begin{equation}
M_\bullet(\tau) = M_\bullet(0)\, e^{\tau/\tau_{\rm ef}}+ \int_0^\tau{\rm d}\tau' e^{-(\tau'-\tau)/\tau_{\rm ef}}\, \dot{M}_{\bullet, \rm DF}(\tau')~.
\label{massacentrale}
\end{equation}
Since $\dot M_{\bullet, \rm acc}$ is proportional to the central BH mass, at early times disk accretion is expected to be subdominant with respect to dynamical friction, and to dominate at late times.

In the left panel of Fig.~\ref{fig:mass} we illustrate the growth of the central supermassive BH, and the contribution from dynamical friction and Eddington accretion ($\lambda=1$) for a galaxy at redshift $z\approx 7.5$ with SFR $\psi\sim 100\, M_\odot$ yr$^{-1}$, apt for the typical hosts and progenitors of the most distant quasars (e.g., Venemans et al. 2017a,b, 2018). The early growth is dominated by dynamical friction, which originates an heavy BH seed with mass $M_\bullet\sim 10^4-10^6\, M_\odot$ within a galactic age $\tau\sim$ some $10^7$ yr. For older ages, disk accretion progressively takes over and leads to accumulate masses $M_\bullet\gtrsim 10^9\, M_\odot$ within some $10^8$ yr. We compare this evolutionary track to the ones for pure disk accretion with $\lambda=1$ and $\lambda=3$. It is seen that to obtain final BH masses of a few $10^9\, M_\odot$ within some $10^8$ yr, super-Eddington accretion with $\lambda\sim 3$ is required if dynamical friction is switched off, while $\lambda=1$ can be retained if dynamical friction enters into the game to build up an heavy seed at early stages; as discussed in Sect.~\ref{intro} this is particularly relevant at $z\gtrsim 7$, where an age of the Universe shorter than $0.8$ Gyr is a demanding constraint. Although a mildly super-Eddington accretion with $\lambda\sim $a few is not implausible at these early cosmic times (e.g., Li 2012; Madau et al. 2014; Aversa et al. 2015; Volonteri et al. 2015; Lupi et al. 2016; Davies et al. 2019; Regan et al. 2019), the formation of an heavy seed by dynamical friction as proposed here may constitute an alternative explanation or a complementary process.

In the right panel of the same Fig.~\ref{fig:mass} we show the mass growth of a BH, in a galaxy at $z\approx 2$ with SFR $\psi\sim 300\, M_\odot$ yr$^{-1}$, representative of a typical ETG progenitor at the peak of the cosmic star formation history and at the knee of the SFR function (e.g., Gruppioni et al. 2013, 2015; Mancuso et al. 2016; Lapi et al. 2017, 2018). In this case the evolutionary tracks with dynamical friction and disk accretion for $\lambda=1$ and $\lambda=0.3$ are compared to that for pure disk accretion with $\lambda=1$. This is to show that, even at these intermediate redshifts, the dynamical friction mechanism allows to effectively create heavy seeds within $\tau\sim$ some $10^7$. As discussed in Sect.~\ref{intro} these may help to attain BH masses in excess of several $10^8-10^9\, M_\odot$ within a time $\tau_\psi\sim$ some $10^8$ yr (the typical duration of the star formation and BH accretion in massive ETG progenitors; see discussion in Sect.~\ref{intro}), even with Eddington ratios $\lambda\sim 0.3$ appreciably smaller than $1$ that are on the average suggested by single-epoch measurements in quasars out to $z\lesssim 4$ (see Vestergaard \& Osmer 2009; Kelly \& Shen 2013; Vestergaard 2019).

\section{Probing the BH seed growth via GW emission}\label{GW}

The contribution to the early BH growth from migration of compact remnants by gaseous dynamical friction in high-$z$ ETG progenitors could hardly be probed via standard electromagnetic observations; even if it were present, luminous emission would be too weak and likely strongly dimmed by the very gas and dust-rich environment to be ever detected. However, we will show that the repeated mergers of the compact remnants with the accumulating central BH mass can originate detectable GW signals (e.g., Barausse 2012; Barack et al. 2019). Specifically, in this Section we aim to compute the cosmic-integrated GW rate density of these events as a function of redshift, and their detectability with the future ET and LISA detectors.

Eq.~(\ref{massacentrale}) establishes a one-to-one correspondence between the galaxy age $\tau$ and the value of the central BH mass $M_\bullet(\tau|\psi,z)$; moreover, the latter can be combined with a given mass $m_\bullet$ of the migrating compact remnant to construct the chirp mass $\mathcal{M}_{\bullet\bullet}=(M_\bullet\, m_\bullet)^{3/5}/(M_\bullet+m_\bullet)^{1/5}$, which determines the strength of the GW signal associated to each merging event. Thus the rate of merging events due to dynamical friction per bin of chirp mass is obtained easily from Eq.~(\ref{rmergerv}) by a change of variable, in the form
\begin{equation}
\frac{{\rm d}\dot{N}_{\rm DF}}{{\rm d}\mathcal{M_{\bullet\bullet}}}(\mathcal{M}_{\bullet\bullet},\tau|\psi,z)=\frac{{\rm d}\dot{N}_{\rm DF}}{{\rm d} m_\bullet}(m_\bullet,\tau|\psi,z)\, \frac{{\rm d} m_\bullet}{{\rm d}\mathcal{M}_{\bullet\bullet}}~.
\end{equation}
We now can compute the cosmic rate density of merging events due to dynamical friction per unit chirp mass $\mathcal{M}_{\bullet\bullet}$ and comoving cosmic volume $V$ as a function of redshift $z$ (or equivalently cosmic time $t_z$) as
\begin{equation}
\begin{aligned}
\frac{{\rm d}\dot{N}_{\rm DF}}{{\rm d} V{\rm d}\mathcal{M}_{\bullet\bullet}}&(\mathcal{M}_{\bullet\bullet},z)=\int{\rm d}\psi\,\frac{{\rm d} N}{{\rm d} V{\rm d}\psi}(\psi,z)\,\int_{t_z-\tau_{\psi}}^{t_z}{\rm d} t_{z_{\rm form}}\times\\
\\
&\times\frac{{\rm d} p}{{\rm d} t_{z_{\rm form}}}(t_{z_{\rm form}}|\psi)\,\frac{{\rm d}\dot{N}_{\rm DF}}{{\rm d}\mathcal{M}_{\bullet\bullet}}(\mathcal{M}_{\bullet\bullet},t_z-t_{z_{\rm form}}|\psi,z)~;
\end{aligned}
\label{mergerratechirp}
\end{equation}
here the quantity ${\rm d}\dot{N}_{\rm DF}/{\rm d}\mathcal{M_{\bullet\bullet}}$ is computed at a galaxy age $\tau=t_z-t_{z_{\rm form}}$, where $t_{z_{\rm form}}$ is the cosmic time at which the galaxy has started its main star formation episode. This expression is then integrated over $t_{z_{\rm form}}$ weighting by the appropriate distribution of formation redshift ${\rm d}p/{\rm d}t_{z_{\rm form}}$, that for the sake of simplicity here we take to be flat between $t_z$ and $t_z-\tau_\psi$, with $\tau_\psi\sim$ some $10^8$ yr being the star formation duration as inferred from the galaxy main sequence (see Rodighiero et al. 2011; Speagle et al. 2014; also discussion in Sects.~\ref{intro} and \ref{accretion}). Finally, an integration over the possible values of the SFR $\psi$ is performed, by weighting with the galaxy SFR functions ${\rm d} N/{\rm d} V\,{\rm d}{\psi}$ at cosmic time $t_z$; these galaxy statistics have been determined observationally over a wide range of SFRs $\psi\sim 10-3000\, M_\odot$ yr$^{-1}$ and redshifts $z\sim 0-8$ thanks the combination of deep UV/near-IR/far-IR/submm/radio surveys (see Mancuso et al. 2016a,b; Lapi et al. 2017; Boco et al. 2019, their Fig.~1).

The outcome of Eq.~(\ref{mergerratechirp}) is illustrated in the left panel of Fig.~\ref{fig:chirpq} at different redshifts (color-coded). The shape of the curves is mainly determined by the evolution of the central BH mass; it grows by continuous merging with the
stellar compact remnants due to gaseous dynamical friction and by disk accretion. In the early stages the dynamical friction process dominates, and the time spent by the central BH in a given (logarithmic) mass bin increases with the BH mass; this in turn originates an increasing behavior of the chirp mass distribution at low $\mathcal{M}_{\bullet\bullet}$. In the late evolution, the disk Eddington-like accretion takes over, and the time spent by the central BH in a given (logarithmic) mass bin is independent of the BH mass; thus the chirp mass distribution flattens at large $\mathcal{M}_{\bullet\bullet}$; the final drop is related to the absence of chirp masses $\mathcal{M}_{\bullet\bullet}\gtrsim 10^5\, M_\odot$ since this extreme value would correspond to the coalescence of a central BH of $M_\bullet\approx 3\times 10^9\, M_\odot$ with a stellar remnant of $m_\bullet\approx 100\, M_\odot$.

Another useful quantity is the probability distribution of mass ratios $q\equiv m_\bullet/M_{\bullet}$ at given chirp mass $\mathcal{M}_{\bullet\bullet}$, averaged over the galaxy population. To this purpose we relate each galactic age $\tau$ to the central BH mass $M_{\bullet}$ via Eq.~(\ref{massacentrale}), and then express both $M_\bullet = \mathcal{M}_{\bullet\bullet}\, q^{-3/5}\,(1+q)^{1/5}$ and the merging compact remnant mass $m_\bullet= \mathcal{M}_{\bullet\bullet}\, q^{2/5}\,(1+q)^{1/5}$ in terms of $\mathcal{M}_{\bullet\bullet}$ and of $q$ themselves. The mass ratio distribution from the dynamical friction process is then given by
\begin{equation}
\begin{aligned}
\frac{{\rm d}p_{\rm DF}}{{\rm d}q}&(q|\mathcal{M}_{\bullet\bullet},z)\propto \int{\rm d}\psi\,\frac{{\rm d} N}{{\rm d} V{\rm d}\psi}(\psi,z)\times\\
\\
&\times\frac{{\rm d}m_\bullet}{{\rm d}q}\, \frac{{\rm d}\dot{N}_{\rm DF}}{{\rm d}m_\bullet}[m_\bullet(\mathcal{M}_{\bullet\bullet},q),\tau(\mathcal{M}_{\bullet\bullet}, q)|\psi,z]
\end{aligned}
\label{qdist}
\end{equation}
with the normalization constant determined by the condition $\int{\rm d}q~ {\rm d}p_{\rm DF}/{\rm d}q=1$. The result is illustrated in the right panel of Fig.~\ref{fig:chirpq} at $z\sim 2$ (it is similar at other redshifts) for different chirp masses (color-coded). The behavior of each curve is easily understood, mirroring the shape of the merging rate (cf. Fig.~\ref{fig:mergerrate}). Specifically, the peak at smaller $q$ refers to the neutron stars while that at larger $q$ to the most common $50\, M_\odot$ compact remnants merging with the central BH; as the mass of the latter increases, the chirp mass raises too, and the mass ratio distribution shifts toward lower $q$ retaining a similar shape. Integrating Eq.~(\ref{mergerratechirp}) over the chirp masses eventually yields the cosmic rate density of merging events due to dynamical friction during BH seed formation
\begin{equation}
\frac{{\rm d}\dot{N}_{\rm DF}}{{\rm d} V}(z)=\int{\rm d}\mathcal{M}_{\bullet\bullet}\frac{{\rm d}\dot{N}_{\rm DF}}{{\rm d}\,V{\rm d}\mathcal{M}_{\bullet\bullet}}(\mathcal{M}_{\bullet\bullet},z)\,;
\end{equation}
the result as a function of redshift is shown in Fig.~\ref{fig:confronto_mergerrate}. The event rate density increases quite rapidly from the local Universe toward high redshift, features a peak around $z\approx 2.5$ and then declines steeply. This result is compared with the analogous quantity for the merging of binary compact remnants (e.g., stellar BH-BH) in galaxies, as computed in Boco et al. (2019). The similarity in the redshift evolution of the event rates for the two processes is easily understood since eventually they are both proportional to the birthrate of compact remnants, which is illustrated in the inset. The overall event rate for the process of BH seed formation by gaseous dynamical friction considered here is generally higher by factors $3-10$ than that for the merging of BH-BH binaries; however, one must caveat that the overall normalization of the latter curve is uncertain since it depends on several assumptions regarding complex processes of stellar astrophysics and binary evolution (e.g., binary fraction, common envelope development/survival, supernova kicks, mass transfers, etc.), and has been actually set by comparison with the current AdvLIGO/Virgo measurements in the local Universe (see Abbott et al. 2019; Boco et al. 2019).

\begin{figure*}
    \centering
    \includegraphics[width=.49\textwidth]{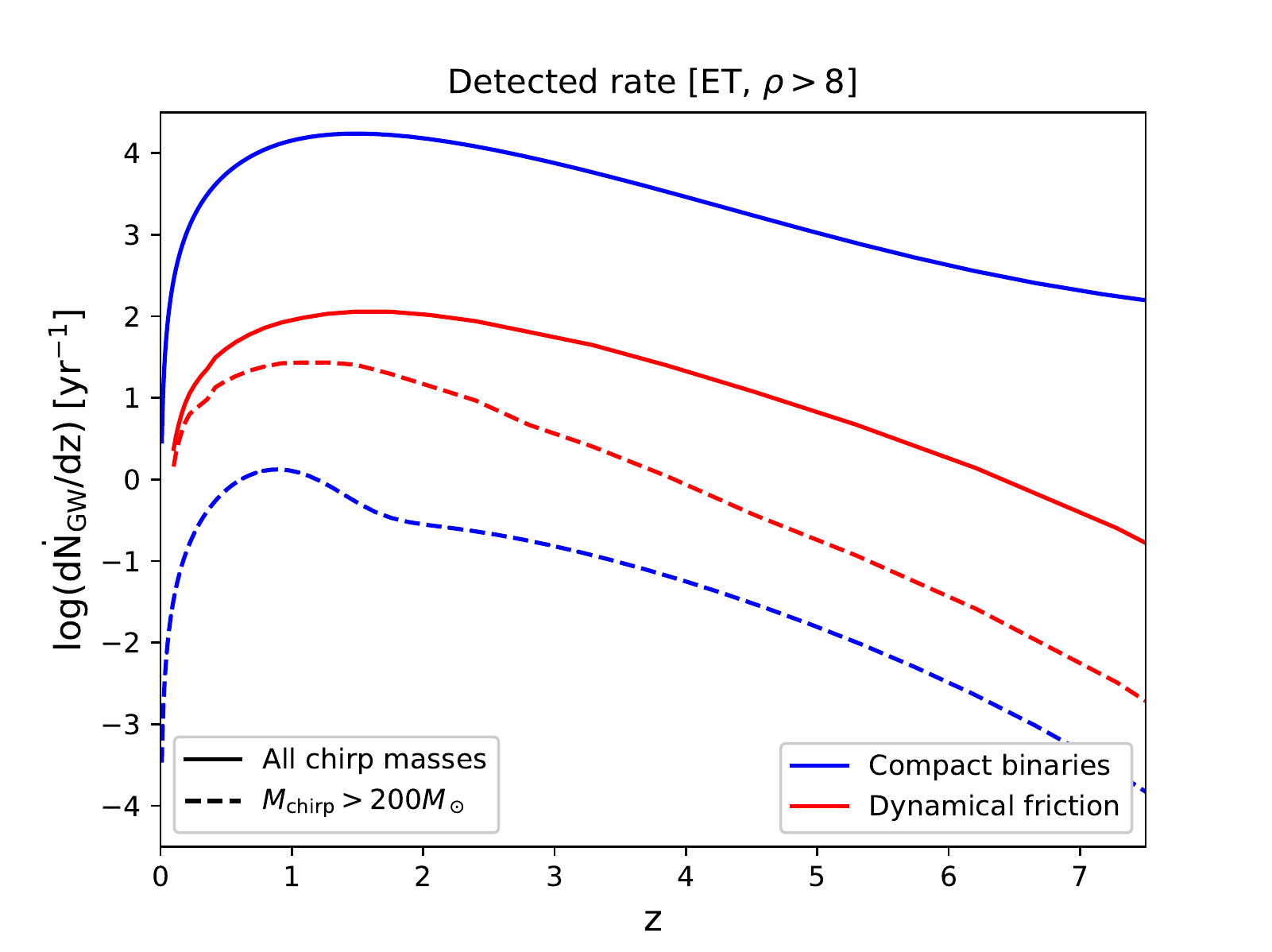}
    \includegraphics[width=.49\textwidth]{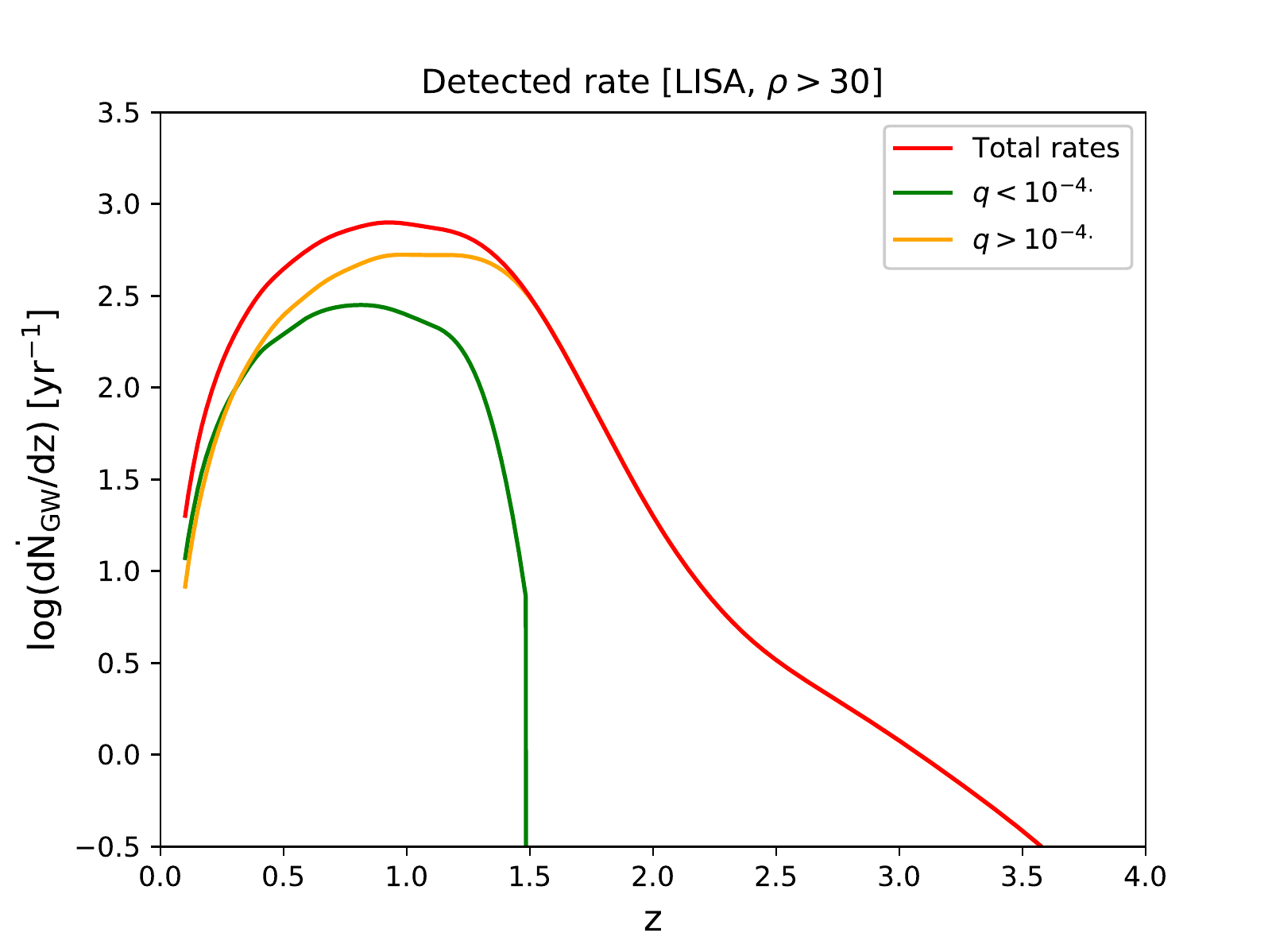}
    \caption{Rate ${\rm d}\dot N_{\rm GW}/{\rm d}z$ of GWs events as a function of redshift detectable by ET (left panel) with a SNR $\rho>8$ and by LISA (right panel) with a SNR $>30$. In the left panel, the red lines refer to the GWs emitted by the merging events due to gaseous dynamical friction, while the blue lines refer to the GW events from merging of compact binaries in galaxies; solid lines include all events irrespective of the chirp mass, while dashed lines is the contribution from events with chirp masses larger than $\mathcal{M}_{\bullet\bullet}>200\, M_\odot$. In the right panel, red line is the total rate of GWs emitted by the merging events due to gaseous dynamical friction, while green and orange lines highlight the contribution from events with mass ratios smaller or largen than $q\sim 10^{-4}$, representative of extreme and intermediate mass ratio inspirals, respectively.}
    \label{fig:detected}
\end{figure*}

\subsection{Rates and properties of detectable GW events}

We now investigate the detectability of the merging events associated to the BH growth via dynamical friction and disk accretion by the future ground and space-based instruments, and in particular the ET (see \texttt{http://www.et-gw.eu/}) and LISA (see \texttt{https://www.elisascience.org/}). Given their diverse frequency sensitivity bands, these detectors provide complementary information; specifically, ET will preferentially pinpoint the early stages of the process when the central BH has still a comparable mass to the migrating stellar remnants, while LISA will probe the subsequent phase when the central BH has already accumulated a mass much larger than that of the remnants, so as to originate intermediate to extreme mass ratio inspirals.

We are interested in estimating the rate of detected events from redshift $z$ with a sky-averaged signal-to-noise ratio (SNR) $\sqrt{\bar{\rho^2}}$ exceeding a given threshold $\rho_0$ (standard values of $\rho_0=8$ for ET and around $30$ for LISA are chosen); this can be written as
\begin{widetext}
\begin{equation}
\begin{aligned}
\frac{{\rm d}\dot{N}_{\rm GW}}{{\rm d} z}(z,>\rho_0) &=\frac{1}{1+z}\,\frac{{\rm d} V}{{\rm d} z}\,\int{\rm d}\mathcal{M}_{\bullet\bullet}~\frac{{\rm d}\dot{N}_{\rm DF}}{{\rm d} V{\rm d}\mathcal{M}_{\bullet\bullet}}(\mathcal{M}_{\bullet\bullet},z)\,\int{\rm d}q\, \frac{{\rm d}p_{\rm DF}}{{\rm d}q}(q|\mathcal{M_{\bullet\bullet}},z)\times\\
\\
&\times \int{\rm d}\Delta t_{\rm obs}\, \frac{{\rm d}p}{{\rm d}\Delta t_{\rm obs}}\,\Theta_{\rm H}\left[\sqrt{\bar{\rho^2}}(\mathcal{M}_{\bullet\bullet}, q, \Delta t_{\rm obs},z)\gtrsim \rho_0\right]~.
\end{aligned}
\label{detected}
\end{equation}
\end{widetext}
In this expression ${\rm d}V/{\rm d}z$ is the comoving volume per unit
redshift interval, the factor $1/(1+z)$ takes into account cosmological time dilation, ${\rm d}\dot N_{\rm DF}/{\rm d}V\,{\rm d}\mathcal{M}_{\bullet\bullet}$
is the cosmic rate density from Eq.~(\ref{mergerratechirp}), ${\rm d}p_{\rm DF}/{\rm d}q$ is the mass ratio distribution from Eq.~(\ref{qdist}), ${\rm d}p/{\rm d}\Delta t_{\rm obs}$ is the probability distribution of observing the event for a time interval $\Delta t_{\rm obs}$ (this is especially relevant for LISA observations, see below), and finally the Heaviside step function $\Theta_{\rm H}[\cdot]$ specifies that only events with sky-averaged SNR $\sqrt{\bar{\rho^2}}$ (that depends on all these variables) in excess of the threshold $\rho_0$ must be considered in the detection rate estimation.

We evaluate the sky-averaged SNR as
\begin{equation}
\begin{aligned}
\sqrt{\bar{\rho^2}} &= 8\, \sqrt{\frac{2}{25}}\, \left(\frac{20}{3}\right)^{5/6}\,\frac{R_0}{D_L(z)}\, \left[\frac{(1+z)\,\mathcal{M}_{\bullet\bullet}}{ M_\odot}\right]^{5/6}\times\\
\\
&\times\zeta_{\rm max}^{1/2}(\mathcal{M}_{\bullet\bullet}, q, \Delta t_{\rm obs}, z)~.
\end{aligned}
\label{SNR}
\end{equation}
In the above $D_L(z)$ is the luminosity distance from the GW source at redshift $z$, while $R_0$ is the detector characteristic distance parameter; this is commonly written as
\begin{equation}
R_0^2 = \frac{25\, M_\odot^2}{192\, \pi\, c^3}\,\left(\frac{3\, G}{20}\right)^{5/3}\, x_{7/3}
\label{R02}
\end{equation}
in terms of the auxiliary quantity
\begin{equation}
x_{7/3} = \frac{1}{(\pi\, M_\odot)^{1/3}}\,\int_0^\infty\,\frac{{\rm d}f}{f^{7/3}\, S(f)}~.
\end{equation}
Here $S(f)=R(f)\, P_n(f)+S_c(f)$ represents the total sensitivity curve, that includes the sky and polarization averaged response function $R(f)$ of the instrument, the instrumental noise $P_n(f)$, and the confusion noise $S_c(f)$. For ground-based detectors like AdvLIGO/Virgo and ET, $S_c(f)$ is usually neglected and $R(f)\simeq 5$ holds independently of the frequency (in some previous works $x_{7/3}$ is defined in terms of $P_n(f)$ and the quantity $1/R(f)\approx 1/5$ is included in the prefactor of Eq.~\ref{R02} defining $R_0^2$). For LISA instead $R(f)$ is a complex frequency dependent function and $S_c(f)$, mainly due to unresolved galactic binaries, must be taken into account. We adopt the sensitivity curves by Hild et al. (2011) for ET and by Robson et al. (2019) for LISA. Finally, coming back to Eq.~(\ref{SNR}) the function
\begin{equation}
\zeta_{\rm max} = \frac{1}{(\pi\, M_\odot)^{1/3}\, x_{7/3}}\, \int_{f_{\rm in}}^{f_{\rm isco}}\,\frac{{\rm d}f}{f^{7/3}\, S(f)}
\end{equation}
specifies the overlap of the signal waveform with the observational bandwidth during the inspiral phase of the event (here we exclude merger and ringdown phases since for intermediate/extreme mass ratio binary mergers their modeling is quite uncertain). The upper limit of integration is taken to be the redshifted GW frequency at the innermost circular stable orbit (in a Schwarzschild spacetime since for simplicity a non-spinning BH is assumed), given by
\begin{equation}
f_{\rm isco} = \frac{1}{6\sqrt{6}\, \pi\, (1+z)}\, \frac{c^3}{G M_{\rm bin}}\approx \frac{4400}{1+z}\, \left(\frac{M_{\rm bin}}{M_\odot}\right)^{-1}\,\,{\rm Hz}
\end{equation}
where $M_{\rm bin} = m_\bullet + M_\bullet = \mathcal{M}_{\bullet\bullet}\, (1+q)^{6/5}\, q^{-3/5}$ is the total mass of the binary (see Finn 1996; Taylor \& Gair 2012). The lower limit of integration is an initial GW frequency $f_{\rm in}$ that takes into account the evolution during the observation time. For ground-based instrument like ET the frequency shift is very rapid and one can approximately take $f_{\rm in}\simeq 0$ in the integral defining $\xi_{\rm max}$, so that the SNR in Eq.~(\ref{SNR}) is independent of $\Delta t_{\rm obs}$ and the distribution ${\rm d}p/{\rm d}\Delta t_{\rm obs}$ in Eq.~(\ref{detected}) integrates to unity and does not matter. Contrariwise, for LISA the frequency evolution is quite slow, and one can determine $f_{\rm in}$ by integrating the orbital averaged equations (see Peters 1964) to obtain
\begin{equation}
f_{\rm in}\simeq  f_{\rm isco}\, \left[1+\frac{1}{5}\,\left(\frac{2}{3}\right)^4\, \frac{q^{8/5}}{(1+q)^{16/5}}\, \frac{c^3\, \Delta t_{\rm obs}}{G\mathcal{M_{\bullet\bullet}}\, (1+z)}\right]^{-3/8}~;
\end{equation}
since $f_{\rm isco}$ can be reached at a random time during the mission, we take ${\rm d}p/{\rm d}\Delta t_{\rm obs}$ as a flat distribution between zero and the nominal mission duration, that for LISA is $4$ yr.

\begin{figure*}
    \centering
    \includegraphics[width=.49\textwidth]{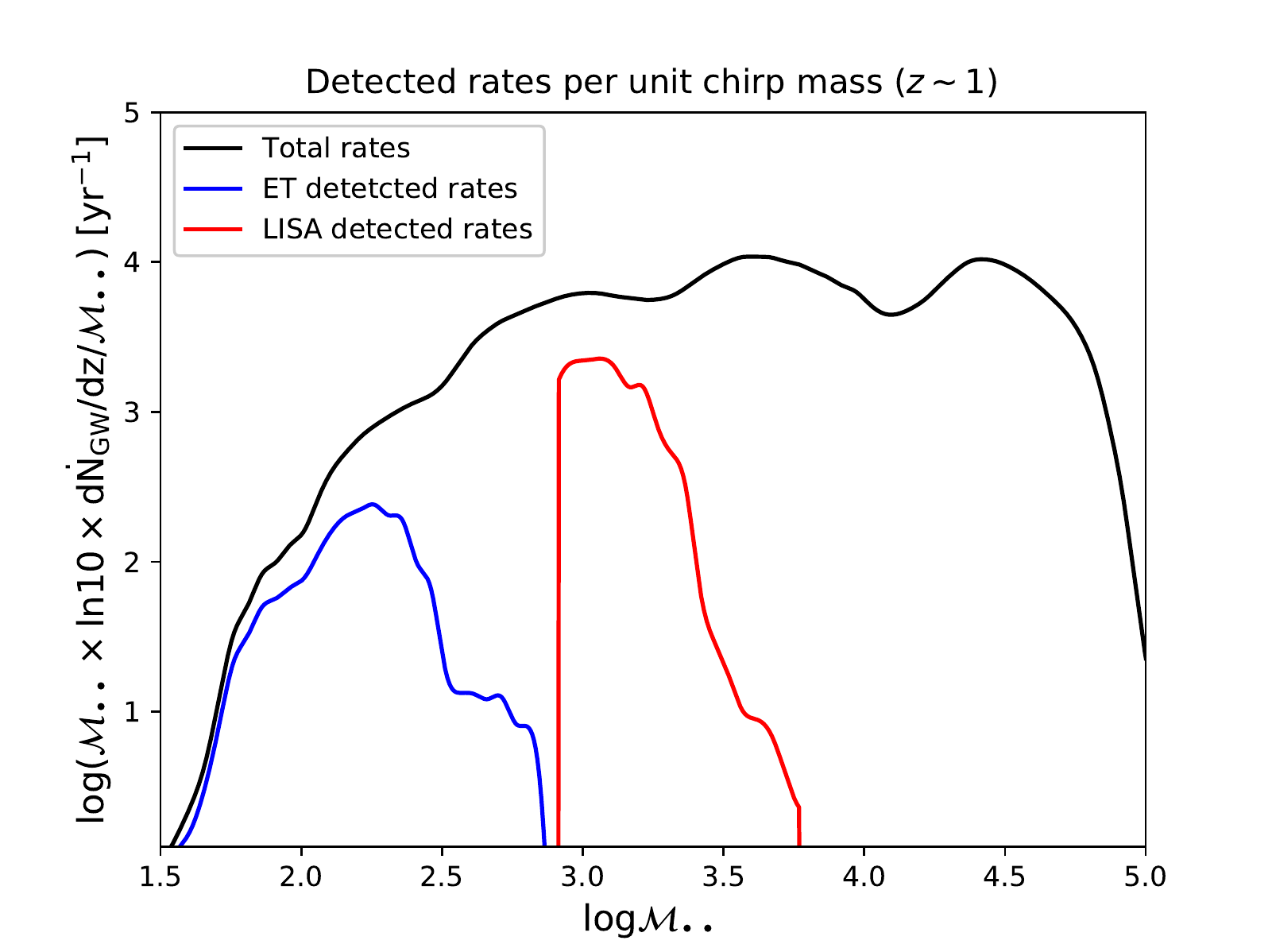}
    \includegraphics[width=.49\textwidth]{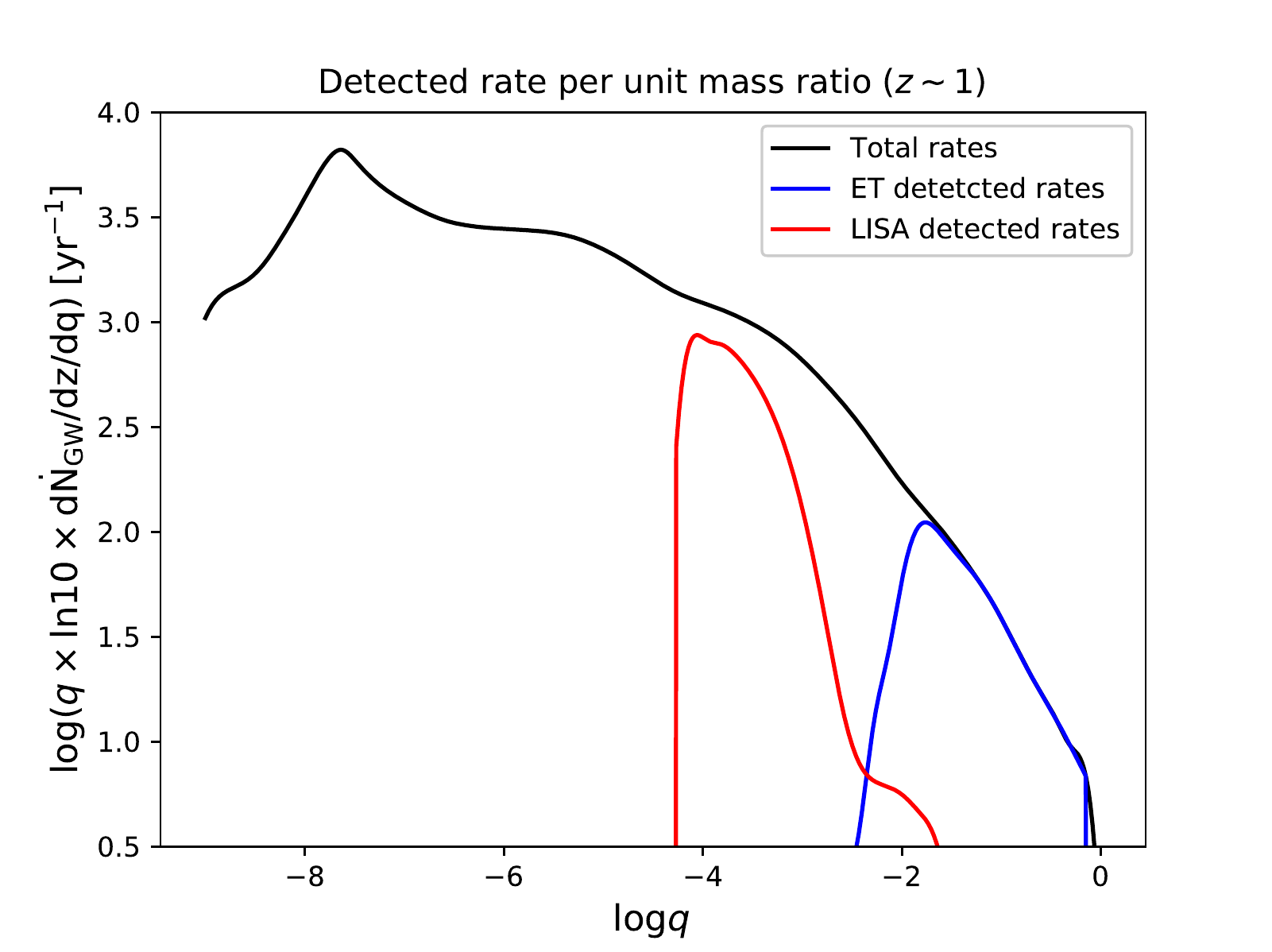}
    \caption{Chirp mass (left panel) and mass ratio (right panel) distributions of the merging events due to gaseous dynamical friction at $z\sim 1$, that can be detected by ET and LISA. Black lines are the intrinsic distributions of the events, while the blue and red lines highlight the portions of them that can be detected by ET with SNR $\rho>8$ and by LISA with SNR $\rho >30$, respectively.}
    \label{fig:detected_confronto}
\end{figure*}

In the left panel of Fig.~\ref{fig:detected} we show the GW event rate as a function of redshift, detectable by ET with a SNR $\rho>8$. The shape reflects the intrinsic cosmic merger rate in Fig.~\ref{fig:confronto_mergerrate}; however, the detected rate is strongly suppressed due to the frequency sensitivity band of ET. In fact, a ground-based interferometer like ET is sensitive only to high-frequency GWs $\gtrsim$ a few Hz), that correspond to merging event with low chirp mass $\mathcal{M}_{\bullet\bullet}\lesssim 500\, M_\odot$; in our context these occur only in the early stages of the evolution when the central BH is still modest, and coalesces with the comparably small masses of the migrating stellar compact remnants. The overall rates detectable by ET amounts to a few hundreds events per yr spread over a wide redshift range $z\sim 0-6$. We stress that the detected rate from the standard merging of compact binaries (color-code) in galaxies overwhelms that from the dynamical friction process considered here by factors $10^2-10^3\, M_\odot$. Nevertheless, the contribution from events with chirp masses larger than a few $10^2\, M_\odot$ is larger for the dynamical friction process, since the number of very high mass compact binaries is strongly suppressed by the stellar initial mass function. Thus recognizing a sizeable number of events with such large $\mathcal{M}_{\bullet\bullet}$ in $z\gtrsim 1$ could be a marking feature of the BH seed formation process by gaseous dynamical friction. Note that a similar computation for AdvLIGO/Virgo (design sensitivity) yields an overall number $\lesssim 10$ events per yr, of which only $\lesssim 0.3$ per yr with chirp mass $\mathcal{M}_{\bullet\bullet}\gtrsim$ a few $10^2\, M_\odot$.

In the right panel of Fig.~\ref{fig:detected} the same is illustrated for LISA with SNR $\rho>30$. Given its sensitivity band, the detected events will correspond to chirp masses $\mathcal{M}_{\bullet\bullet}\sim 500-5000\, M_\odot$; in our context these originate at late galactic ages from the merging of stellar compact remnants with an already large central BH mass $M_\bullet\sim 10^4-10^6\, M_\odot$. The detected rate of these intermediate and extreme mass ratio inspirals peaks around $z\sim 1$ and then declines steeply because the SNR falls below threshold; the color-code highlights the different behavior for intermediate ($q\gtrsim 10^{-4}$) and extreme ($q\lesssim 10^{-4}$) mass ratio inspirals, with the latter providing an appreciable contribution only at $z\lesssim 1$.

Finally, the complementarity of ET and LISA is better highlighted in Fig.~\ref{fig:detected_confronto}, where the rate per unit chirp mass and the rate per unit mass ratio of the events detected at $z\sim 1$ by the two detectors (color-coded) are compared with the intrinsic ones; it is seen that ET mostly probes the events with $\mathcal{M}_{\bullet\bullet}\sim 50-500\, M_\odot$ and $q\sim 0.01-1$, while LISA probes the rate at $\mathcal{M}_{\bullet\bullet}\sim 500-5000\, M_\odot$ and $q\sim 10^{-5}-10^{-2}$.

\section{Discussion}\label{discussion}

In this Section we aim to critically discuss some of the main assumptions underlying our (semi)analytic treatment, that may affect the values of the dynamical friction timescale and its dependence on the physical parameters of the gaseous environment and of the migrating compact remnants. Although a full assessment of these effects is beyond the scope of the present paper, we provide here some  order-of-magnitude estimates that could help the reader to understand the present limitations of our work.

\begin{itemize}
    \item \emph{Large-scale clumpiness.} We have assumed a smooth density distribution of the inner star-forming gas. Actually, the structure of high-$z$ ETG starforming progenitors is more complex. On kpc scales, both observations (e.g., Genzel et al. 2011; Tadaki et al. 2017a,b, 2018; Hodge et al. 2019; Lang et al. 2019; Rujopakarn et al. 2019) and simulations (e.g., Bournaud et al. 2014; Mandelker et al. 2014, 2017; Oklopcic et al. 2017) indicate the presence of clumps with masses $10^7-10^8\, M_\odot$ and sizes of $100-200$ kpc; note that even more massive and extended clumps can be present but are rarer, and could be real outcomes from collisions of smaller ones (e.g., Tamburello et al. 2015) or apparent structures due to blending from observations with limited resolution (e.g., Tamburello et al. 2017; Behrend et al 2016; Faure et al. 2019, in preparation). The survival of the clumps is still a debated issue, with different simulations favoring short-lived clumps because of feedback and/or collisions (e.g., Hopkins et al. 2012; Oklopcic et al. 2017), or long-lived clumps that may eventually sink toward the center via gravitational torque and bar instabilities and contribute to the growth of a central bulge (e.g., Ceverino et al. 2012; Bournaud et al. 2014). Indubitably, the presence of such a clumpiness in the gaseous medium may in principle affect the dynamical evolution of the remnants. However, high-resolution observations with ALMA (see Hodge et al. 2019; Rujopakarn et al. 2019) have revealed that such clumps contribute less than $10\%$ of the overall star-formation; the latter mainly occurs in a rather smooth gaseous and dust-enshrouded medium within the central kpc scale. Provided that in our treatment most of the compact remnants effectively contributing to the growth of the central BH seed come from initial radii of $\lesssim 300$ pc, the assumption of a smooth distribution for the inner star-forming gas should hold to a good approximation.

    \item \emph{Molecular clouds and stellar clusters.} On sub-kpc scales star formation is likely to occur preferentially in molecular gas clouds with masses $10^6\, M_\odot$ and radii of $10-20$ pc. Observations show a rather smooth distribution of the stellar mass in high-$z$ star-forming systems (e.g., Swinbank et al. 2010; Hodge et al. 2016; Rujopakarn et al. 2016; Lang et al. 2019) and in their quiescent high-$z$ (e.g., van der Wel \& van der Marel 2008; Belli et al. 2017) and local descendants (e.g., Cappellari et al. 2013), indicating that molecular clouds are dissolved or a substantial amount of stars can escape quite rapidly from them (typical escape timescales amount to $\lesssim 100$ Myr). However, it could be that some compact remnants born within the cloud might remain bound to a stellar cluster originated there; this will reduce somewhat the number of remnants available for growing the central BH seed. On the other hand, during the formation of the bulge, the stellar clusters may themselves migrate toward the central region via dynamical friction against the background stars, and contribute to the growth of a nuclear star-cluster there (e.g., Antonini et al. 2015).

    \item \emph{Local feedback from SN explosions.} Some progenitors of the compact remnants can have undergone a SN explosion, possibly removing a sizeable amount of gas from its surroundings; this in principle could hamper the effectiveness of the gaseous dynamical friction process. However, two occurrences mitigate the effect. First, most of remnants relevant to the growth of the BH seed are formed and migrate to the center within a few $10^7$ yr, when the average gas metallicity amounts to $\lesssim Z_\odot/10$ (e.g., Pantoni et al. 2019; Boco et al. 2019); in these conditions, most of the remnants are formed by direct collapse without undergoing a SN explosions (e.g., Spera et al. 2015; Spera \& Mapelli 2017). Second, even if the SN explodes, it can efficiently sweep up material during the  energy-conserving expansion phase, out to a radius $R_{\rm SN}\sim 5\, t_4^{2/(5-\alpha)}\,n_{2}^{-1/(5-\alpha)}\, E_{51}^{1/(5-\alpha)}$ pc where $E_{51}\equiv E_{\rm SN}/10^{51}$ erg is the energy of a SN explosion, $n_2\equiv n/10^2$ cm$^{-3}$ is the average gas density and $t_4\equiv t/10^4$ yr the time since the explosion (e.g., Ostriker \& McKee 1988; Mo et al. 2010); however, once formed the remnant will move in the gaseous medium at a typical velocity of $\sigma_{200}\equiv \sigma/200$ km s$^{-1}$ and thus will travel a distance $R_{\rm rem}\sim 2\, \sigma_{200}\,t_4$ pc, implying that most of the gas mass swept up by the remnant is replaced after $\lesssim 10^5$ yr.

    \item \emph{Feedback from the central BH seed.} Though at early times the BH seed growth is dominated by dynamical friction, some gas accretion onto it can occur; the ensuing feedback can partially remove gas from the central region, so offsetting further migration of compact remnants by dynamical friction. The timescale for gas evacuation out to a radius $R_{100}\equiv R/100$ pc from the center due to a BH momentum-driven wind can be estimated as $t_{\rm evac}\sim 8\times 10^7\, R_{100}\, \sigma_{200}\, M_{\bullet,4}^{-1/2}$ yr (see King 2003; King \& Pounds 2015), where $\sigma_{200}\equiv \sigma/200$ km s$^{-1}$ is the galaxy velocity dispersion and $M_{\bullet,4}\equiv M_{\bullet}/10^4\, M_\odot$ is the BH mass; however, the dynamical time for the gas to refill such a region amounts to $t_{\rm dyn}\sim 5\times 10^5\, R_{100}\, \sigma_{200}^{-1}$ yr. Thus the feedback from the central BH will become truly effective as its mass attains $M_\bullet\sim 2.5\times 10^8\, M_\odot\, \sigma_{200}^4$, when however the growth is already largely dominated by gas accretion. Incidentally, note that this condition has also been invoked to explain the $M_\bullet-\sigma$ relation observed between the relic supermassive BH mass and the velocity dispersion of the old population in ETGs (e.g., King \& Pounds 2015; Kormendy \& Ho 2013; McConnell \& Ma 2013; Shankar et al. 2016).

    \item \emph{Three-body encounters.} Interactions among the central BH seed and two migrating remnants, that can eject from the central region the least massive one (e.g., Hills \& Fullerton 1980), could in principle reduce the efficiency of the dynamical friction process in growing the seed. A detailed assessment of the issue clearly require a full dynamical simulation, but a simple argument can be provided along the following lines. The typical radius $r_\bullet$ within which the migrating stellar remnants start to feel the dynamical influence of the central BH seed with mass $M_{\bullet,4}\equiv M_\bullet/10^4\, M_\odot$ can be computed as $r_\bullet\sim G\, M_\bullet/\sigma^2(r_\bullet)$; on considering the approximate scaling with radius $r_{\rm pc}\equiv r/$pc of the velocity dispersion $\sigma(r)\approx 80\, r_{\rm pc}^{\alpha/2}$ km s$^{-1}$, this implies $r_{\bullet}\approx 0.05\, M_{\bullet,4}^{1/(1+\alpha)}$ pc. In addition, the timescale for three-body encounters (e.g., Heggie 1975; Binney \& Tremaine 1987; Davies 2002) between the central seed and two remnants of total mass $m_{\bullet,40}\equiv m_\bullet/40\, M_\odot$ can be estimated as $\tau_{\rm 3b}\sim \sigma(r)/2\pi G\, M_\bullet\, \bar n_\bullet(r)\, r\approx 4\times 10^7\, m_{\bullet,40}\,M_{\bullet,4}^{-(4-\alpha)/2\,(1+\alpha)}\, (r/r_\bullet)^{-(2-3\,\alpha)/2}$ yr; here we have evaluated
    $\bar n_\bullet(r)\approx 0.01\, m_{\bullet,40}^{-1}\, r_{\rm pc}^{-\alpha}$ pc$^{-3}$ as the average density of remnants inside the radius $r$ after a galactic age of $\sim 10^7$ yr by taking into account the radial dependence of the gas mass and the fraction of remnants per unit stellar mass according to a Chabrier IMF. The three-body timescale is to be compared with the typical gaseous dynamical friction timescale, that from Eq.~(\ref{taudynamical}) amounts to $\tau_{\rm DF}\approx 10^4\, m_{\bullet,40}^{-1}\, M_{\bullet,4}^{5/2\,(1+\alpha)}\,(j/j_c)^{3/2}\, (r_c/r_\bullet)^{5/2}$ yr; thus whenever $r_c$ is close to $r_\bullet$, as required to have effective three-body interactions, $\tau_{\rm DF}\ll \tau_{\rm 3b}$ holds so that dynamical friction is expected to wash out the dynamical effects of possible three-body encounters.

    \item \emph{Velocity structure.} We have adopted a velocity structure dominated by random motions with velocity dispersion $\sigma(r)$, that has been computed via the Jeans equation taking into account the overall mass distribution. Actually the situation in star-forming ETG progenitors is slightly more complex. On a scale of a few kpc, the velocity structure is dominated by rotational motions with $v/\sigma\gtrsim$ a few, in the way of a clumpy unstable disk (see Genzel et al. 2011; Tadaki et al. 2017a,b, 2018; Hodge et al. 2019). However, on sub-kpc scales both observations (e.g., Barro et al. 2016; Rujopakarn et al. 2019) and simulations (e.g., Danovich et al. 2015; Zolotov et al. 2015; Zavala et al. 2016) indicate that dynamical friction, gravitational torques, and violent relaxation will operate toward converting such rotational into random motions, setting up a bulge-like structure with $v/\sigma\lesssim 1$ (see also Lapi et al. 2018). Provided that the majority of the compact remnants contributing to the growth of the central seed BH come from a scale of $\lesssim 300$ pc, our assumption of a dispersion-dominated velocity structure should hold to a good approximation. However, it is still possible that some of the remnants still possess a residual rotational velocity component; in this case the angular momentum increases $j\propto \sqrt{v^2+\sigma^2}$ and the timescale for dynamical friction is correspondingly enhanced as $\tau_{\rm DF}\propto j^{1.5}$. For example, in the rather extreme case $v/\sigma\sim 1$ this amounts to roughly doubling $\tau_{\rm DF}$.

    \item \emph{Dust component.} The central kpc regions of a star-forming ETG progenitors are very dusty (e.g., Tadaki et al. 2017a,b, 2018; Hodge et al. 2019; Rujopakarn et al 2019). In principle, dust can cooperate with the gas component in making the dynamical friction of the compact remnants more efficient, and speed up the building up of the central BH seed. However, quantitatively the effect is expected to be small since the estimated dust-to-gas ratios amount to $M_{\rm dust}/M_{\rm gas}\sim 1-5\%$ (e.g., Berta et al. 2016; Scoville et al. 2016, 2017; Tacconi et al. 2018).

\end{itemize}

We also  warn  that  the  values  of  the  dynamical  friction  timescale  and  its  dependence  on  the physical parameters of the gaseous environment and of the migrating compact remnants could be influenced by other concomitant and co-spatial astrophysical and dynamical processes, not included in our (semi)analytic, orbit-averaged treatment, like natal kicks imparted to the remnants,  stellar  hardening,  development  of  a  circum-binary  disk  around  the  central  BH, tidal stripping effects. It will be most welcome to further investigate the above details via a full hydro+dynamical simulation at high spatial resolution.

\section{Summary and future prospects}\label{summary}

We have proposed a new mechanism for the growth of supermassive BH seeds in the star-forming progenitors of local massive ETGs at $z\gtrsim 1$, that envisages the migration and merging of stellar compact remnants (neutron stars and stellar-mass BHs) via gaseous dynamical friction toward the central high-density regions of such galaxies (see Fig.~1). Our main findings are the following:

\begin{itemize}

\item We have estimated the gaseous dynamical friction timescales in the orbit-averaged approximation, finding that it can be appreciably smaller than $\lesssim 1$ Gyr
for reasonable assumptions on the gas density profile and on the initial conditions of the stellar compact remnants in real and velocity space (see Sect.~\ref{dynamical efficiency} and Table~\ref{tabella2}). We have also provided a fitting formula for the dynamical friction timescale, dependent on the properties of the stellar compact remnants and of the galactic environment (see Eq.~\ref{taudynamical} and Table~\ref{tabella1}).

\item  We have shown that such a process can build up central BH masses of order $10^4-10^6\, M_\odot$ within some $10^7$ yr, so effectively providing heavy seeds before standard disk (Eddington-like) accretion becomes the dominant process for further BH growth toward the locally observed values (see Sect.~\ref{central mass} and Fig.~\ref{fig:mass}). Remarkably, such a process may provide an explanation, alternative to super-Eddington accretion rates, for the buildup of billion solar masses BHs in quasars at $z\gtrsim 7$, when the age of the Universe $\lesssim 0.8$ Gyr constitutes a demanding constraint. Moreover, in more common ETG progenitors at redshift $z\sim 2-6$ it can concur with disk accretion to build such large BH masses even at moderate Eddington ratios $\lesssim 0.3$ within the short star-formation duration $\lesssim$ Gyr of these systems.

\item We have investigated the perspectives to detect the merger events between the migrating stellar compact remnants and the accumulating central supermassive BH via GW emission with future ground and space-based detectors such as ET and LISA (see Sect.~\ref{GW} and Fig.~\ref{fig:detected}). We have computed the redshift distribution of the detectable events, finding that ET will detect at SNR $\gtrsim 8$ a few hundreds events per yr spread over a wide redshift range $z\sim 0-6$, while LISA will reveal at SNR $\gtrsim 30$ several hundreds events in the redshift range $z\sim 0-2$.

\item We have highlighted that ET and LISA can play a complementary role in probing the seed formation process proposed here (see Fig.~\ref{fig:detected_confronto}). In particular, ET will pinpoint the events with chirp masses $\mathcal{M}_{\bullet\bullet}\sim 50-500\, M_\odot$ and mass ratios $q\sim 0.01-1$ occurring at early galactic ages when the stellar remnants can merge with the still comparably small $\sim 10^2-10^3\, M_\odot$ central BH mass; on the other hand, LISA will detect the intermediate and extreme mass ratio inspirals with chirp masses $\mathcal{M}_{\bullet\bullet}\sim 500-5000\, M_\odot$ and $q\sim 10^{-5}-10^{-2}$ occurring at later galactic ages when the stellar compact remnants merge with an already massive $\sim 10^4-10^6\, M_\odot$ central BH.

\end{itemize}

Given the promising findings of this pilot study, it will be most welcome to further investigate the details of the gaseous dynamical friction process, proposed here to grow heavy BH seeds, via a full hydro$+$dynamical simulation at high spatial resolution. In particular, in  Sect.~\ref{discussion}  we have discussed how the dynamical friction timescale and its dependence on the physical parameters of the gaseous environment and of the migrating compact remnants could be influenced by other concomitant and co-spatial astrophysical and dynamical processes, not included in our (semi)analytic, orbit-averaged treatment.

Finally, some future developments or spin-offs of the present work that are worth to be addressed in the near future will include the following: migration of stellar clusters and long-living stars that, if not disrupted or strongly ablated by tidal forces during the orbital decay toward the nuclear regions, may lead to form the nuclear star clusters observed in some local galaxies; occasional formation of multiple seed BHs with masses $\gtrsim 10^3\, M_\odot$ in the central regions, that could eventually merge and produce GW emission detectable by LISA; impact of galaxy- or cluster-scale gravitational lensing on the detected GW event rate presented in this work; reconstruction of the overall BH mass function, from stellar-mass BHs, to transient intermediate-mass BHs (the heavy seeds considered in this study), to supermassive BHs via a continuity equation approach.

\begin{acknowledgements}
We thank the anonymous referee for helpful and constructive comments. We acknowledge A. Bressan, C. Baccigalupi, E. Barausse, A. Macci\'o, and M. Spera for helpful discussions. This work has been partially supported by PRIN MIUR 2017 prot. 20173ML3WW 002 `Opening the ALMA window on the cosmic evolution of gas, stars and supermassive black holes'. AL acknowledges the EU H2020-MSCA-ITN-2019 Project 860744 `BiD4BEST: Big Data applications for Black Hole Evolution STudies', and the MIUR grant `Finanziamento annuale individuale attivit\'a base di ricerca'.
\end{acknowledgements}

\end{document}